\documentstyle[epsf,epsfig,rotate]{aa}

\newcommand{\be}{\begin{equation}}
\newcommand{\ee}{\end{equation}}

\newcommand{\icar}{{\it Icarus, }}

\newcommand{\ana}{{\it A\&A, }}

\newcounter{pp3}
\addtocounter{pp3}{3}

\begin{document}
%\thesaurus{02(02.01.2; 02.13.2; 08.09.2 AA Tau; 08.13.1; 08.16.5)}

\title
{On the Orbital Evolution of Low Mass Protoplanets in 
Turbulent, Magnetised Disks}

\author{Richard P. Nelson}

\institute{ Astronomy Unit, 
 Queen Mary, University of London, Mile End Rd, London, E1 4NS, U.K.}
 
\offprints{R.P.Nelson@qmul.ac.uk}

\date{Received /Accepted}
% \pagerange{\pageref{firstpage}--\pageref{lastpage}}

\def\LaTeX{L\kern-.36em\raise.3ex\hbox{a}\kern-.15em
         T\kern-.1667em\lower.7ex\hbox{E}\kern-.125emX}

%\newtheorem{theorem}{Theorem}[section]

%\label{firstpage}

\titlerunning{Low mass planets in turbulent disks}
\authorrunning{R.P.Nelson }

\abstract{
We present the results of MHD simulations of low mass protoplanets
interacting with turbulent, magnetised protostellar disks. We calculate the
orbital evolution of `planetesimals' and protoplanets with masses 
in the range $0 \le m_p \le 30$ M$_{\oplus}$. The disk models are cylindrical
models with toroidal net-flux magnetic
fields, having aspect ratio $H/r=0.07$ and
effective viscous stress parameter $\alpha \simeq 5 \times 10^{-3}$.

A significant result is that the $m_p=0$ `planetesimals', and protoplanets of 
all masses considered, undergo stochastic migration due to gravitational
interaction with turbulent density fluctuations in the disk.
For simulation run times currently feasible
(covering between 100 -- 150 planet orbits),
the stochastic migration dominates over type I migration for many models.
Fourier analysis of the torques experienced by protoplanets
indicates that the torque fluctuations contain components with significant
power whose time scales of variation are similar to the simulation run
times. These long term torque fluctuations in part explain the
dominance of stochastic torques in the models, and
may provide a powerful means of counteracting the effects of
type I migration acting on some planets in turbulent disks.
The effect of superposing type I migration torques appropriate for laminar
disks on the stochastic torques was examined. This analysis predicts
that a greater degree of inward migration should occur than 
was observed in the MHD simulations.
This may be a first hint that type I
torques are modified in a turbulent disk, but the results are not
conclusive on this matter.

The turbulence is found
to be a significant source of eccentricity driving, with the
`planetesimals' attaining eccentricities in the range $0.02 \le e \le 0.14$
during the simulations. The eccentricity evolution of the protoplanets
shows strong dependence on the protoplanet mass. Protoplanets with
mass $m_p=1$ M$_{\oplus}$ attained eccentricities in the range $0.02 \le e \le
0.08$. Those with $m_p=10$ M$_{\oplus}$ reached $0.02 \le e \le 0.03$.
This trend is in basic agreement with a model in which eccentricity growth
arises because of turbulent forcing, and eccentricity damping occurs through
interaction with disk material at coorbital Lindblad resonances.

These results are significant for the theory of planet formation.
Stochastic migration may provide a means
of preventing at least some planetary cores from migrating into the
central star due to type I migration before they become gas giants.
The growth of planetary cores may be enhanced by preventing
isolation during planetesimal accretion. The excitation of
eccentricity by the turbulence, however, may act to reduce 
the growth rates of planetary cores during the runaway and
oligarchic growth stages, and may cause collisions between planetesimals
to be destructive rather than accumulative.}

\maketitle


\section{Introduction} \label{intro}

The continuing discovery of extrasolar planets by the radial velocity and
transit techniques has generated
renewed interest in the theory of planet formation
(e.g. Mayor \& Queloz 1995;
Marcy, Cochran, \& Mayor 2000; Vogt et al. 2002; Santos et al. 2003).
In the so--called
core instability model of planet formation, gas giant planets form through the
build--up of a rocky and icy core of $\sim 15$ Earth masses, which then
undergoes gas accretion resulting in a giant planet
(e.g. Bodenheimer \& Pollack 1986;
Pollack et al. 1996). The gas accretion stage of this process is
normally believed to take a few million years, during which time
the planetary cores slowly accrete gas through quasi--static
settling, until the planet mass is around 50 Earth masses. Beyond this
mass the gas accretion rate increases dramatically and gas is able to 
accrete onto the planet through a circumplanetary disk
at the rate supplied by the protostellar
disk (Papaloizou \& Nelson 2005). 
An alternative model for giant planet formation is that gravitational
fragmentation of a fairly massive protostellar disk can form
Jovian mass planets directly (e.g. Boss 2001). This model, however,
is unable to account for terrestrial planets, and is unlikely to
explain the existence of Uranus and Neptune.

Since giant planets form in a gas--rich protoplanetary disk, the gravitational
interaction between planets and disks can play an important
evolutionary role.
This interaction
has been the subject of many
studies over the last couple of decades.
In the usual picture, a protoplanet exerts torques on a protostellar disk
through the excitation of spiral density waves at Lindblad resonances,
and possibly through interaction at corotation resonance
(e.g. Goldreich \& Tremaine 1979, 1980; Lin \& Papaloizou 1979;
Papaloizou \& Lin 1984; Ward 1986, 1997;
Tanaka, Takeuchi \& Ward 2002).
The spiral waves carry an associated angular momentum flux,
which is deposited in the disk material
where the waves damp,
leading to exchange of angular momentum
between protoplanet and disk.
The disk lying exterior to the protoplanet orbit exerts a negative 
torque on the planet, and the interior disk exerts a positive torque.
The negative outer disk torque usually dominates and the protoplanet
migrates inward. This is referred to as type I migration.
For protoplanets of $\sim 15$ Earth masses, the migration
time is between $10^4$ and $10^5$ yr (Tanaka, Takeuchi \&
Ward 2002), which is much
shorter than the estimated gas accretion phase of giant
planet formation (Pollack et al. 1996).  Taken at face value,
this presents a serious problem for the core--instability model of
gas giant planet formation. In a recent paper, Papaloizou \& Nelson (2005)
examined the possibility of shortening the formation time scale
of gas giant planets by reducing the dust opacity in the outer
radiative zone of the forming planets. Although the formation time
can be shortened significantly, they concluded that the type I
migration time is always shorter than the planet formation time.
This analysis, however, pertains only to
smooth, laminar disk models.

%For protoplanets in the Jovian mass range, the interaction is non linear
%and gap formation occurs
%(Papaloizou \& Lin 1984; Bryden et al. 1999; Kley 1999). In this case the
%orbital migration of the planet becomes locked to the viscous evolution of
%the disk, and migration is expected to occur on a time scale
%of $10^5$ yr (Lin \& Papaloizou 1986; Nelson et al. 2000; D'Angelo, Kley \&
%Henning 2002). This is usually referred to as type II migration.

Until quite recently most models of viscous accretion disks used the
Shakura \& Sunyaev (1973) $\alpha$ model for the anomalous
disk viscosity. This assumes
that the viscous stress is proportional to the thermal pressure in the disk,
without specifying the origin of the viscous stress.
Balbus \& Hawley (1991) showed that
significant angular momentum transport in weakly magnetised disks could arise
from the magnetorotational instability (MRI).
Non linear numerical simulations performed using a local
shearing box formalism
(e.g. Hawley \& Balbus 1991;
Hawley, Gammie, \& Balbus 1996; Brandenburg et al. 1996)  confirmed
this and showed that the saturated non linear outcome of the MRI
is MHD turbulence
having an effective viscous stress parameter $\alpha$ of between $\sim 5 \times
10^{-3}$ and $\sim 0.1$, depending on the magnetic field configuration.
Global simulations of MHD turbulent disks
(e.g. Armitage 1998; Hawley 2000;
Hawley 2001; Steinacker \& Papaloizou 2002; Papaloizou \& Nelson 2003)
confirm the picture
provided by the local shearing box simulations.

An obvious question is how the interaction of forming planets
with protoplanetary disks changes when the disks are turbulent.
Papaloizou \& Nelson (2003 -- hereafter PN2003)
examined and characterised the turbulence obtained in a variety
of MHD disk models. Nelson \& Papaloizou (2003 -- hereafter NP2003) 
examined the
interaction between a global cylindrical disk model and a massive (5 Jupiter
masses) protoplanet. A similar study was undertaken by
Winters, Balbus, \& Hawley (2003). 
Papaloizou, Nelson, \& Snellgrove (2004 -- hereafter PNS2004)
performed global cylindrical disk simulations
and local shearing box simulations of turbulent disks interacting with
protoplanets of different mass. The main focus of that paper was to
characterise the changes in flow morphology and disk structure as
a function of planet mass, and to examine the transition from
linear to non linear interaction leading to gap formation. 
Nelson \& Papaloizou (2004 -- hereafter NP2004) examined the migration
of low and high mass protoplanets in turbulent disks by examining the
time evolution of the torques exerted on the planets by the disks.
They noticed that low mass planets experienced strongly varying torques
due to interaction with the turbulent density fluctuations, and suggested
that such planets would undergo stochastic migration rather than 
monotonic inward migration normally associated with type I migration
in laminar disks. The planets, however, were maintained on fixed 
circular orbits and so their migration could not be examined
directly.

In this paper we present the results of simulations of low mass
protoplanets embedded in turbulent, magnetised disks, and allow the
planet orbits to evolve due to interaction with the disk.
In order to sample the range of outcomes, we consider a small
ensemble of planets in each simulation.
We find that in all simulations performed,
the forces experienced by the protoplanet
are highly variable, and as a result the protoplanet orbits evolve
similarly to a random walk. A simple analysis suggests that the heavier planets
we consider ought to undergo inward migration predominantly due to the
underlying type I torques, but the simulations on the whole are
not in agreement with this prediction.
Fourier analysis of the torques experienced by the 
planets indicates that they experience stochastic forces with a broad
range of associated time scales of variation, ranging from the planet orbital
period to the simulation run time itself. These long time scale variations
clearly contribute to the long term stochastic behaviour of the planets
observed in the simulations. There is some evidence that
the underlying type I migration is modified in turbulent
disks, but this is not conclusive.

The results also indicate that MHD turbulence is able to drive
significant growth of eccentricities for low mass objects.
In particular, planetesimals and planets with masses similar to the Earth 
can obtain eccentricities $e \sim 0.1$. This clearly has potentially
important consequences for planetary accumulation models. Heavier planets
attain lower eccentricities apparently because their interaction with
the disk at coorbital Lindblad resonances causes eccentricity damping
(e.g. Artymowicz 1993; Papaloizou \& Larwood 2000).

The plan of the paper is as follows. In section~\ref{eqns}
we describe the governing equations.
In sections~\ref{num_method} and \ref{units} we describe the
numerical method and the system
of units used. The initial and boundary conditions used
in the simulations are described in section~\ref{Init-bound},
and the simulation results are presented in 
section~\ref{results}. We discuss the simulation results in 
sections~\ref{stochastic_torques}, \ref{eccentricity}, and \ref{discussion},
focusing in particular on the issues of the balance
between type I and stochastic migration, and the eccentricity evolution of
planets and planetesimals. 
We summarise the paper and draw conclusions in section~\ref{summary}.

\section{Equations of Motion} \label{eqns} 
The equations of ideal MHD written in a frame rotating
with uniform  angular velocity $\Omega_f {\bf {\hat z}} $,
with $ {\bf {\hat z}} $ being the unit vector along the rotation axis
assumed to be  in the vertical direction,
are the continuity equation
\begin{equation}
\frac{\partial \rho}{\partial t}+ \nabla \cdot {\rho\bf v}=0, \label{cont}
\end{equation}
the equation of motion
\begin{equation}
\frac{\partial {\bf v}}{\partial t}
 + {\bf v}\cdot\nabla{\bf v} + 2\Omega_f {\bf {\hat z}}{\bf \times}{\bf v} =
-{\nabla p \over \rho}  -\nabla\Phi +
{(\nabla \times {\bf B}) \times {\bf B}\over 4\pi\rho} \label{mot}
\end{equation}
and the induction equation
\begin{equation}
\frac{\partial {\bf B}}{\partial t}=\nabla \times ({\bf v} \times {\bf B})
\label{induct}
\end{equation}
where ${\bf v}$, $P$, $\rho$, and ${\bf B}$ denote the fluid
velocity, pressure, density  and  magnetic field respectively.
The potential $\Phi = \Phi_{rot} +\Phi_G$
contains contributions due to gravity, $\Phi_G$,
and the centrifugal potential $\Phi_{rot} =
-(1/2)\Omega_f^2|{\bf {\hat z}}\times
{\bf r}|^2.$

The gravitational potential has contributions from a central mass
$M_*$
and $N$ planets with masses $m_{pi}.$
Thus in cylindrical coordinates $(r, \phi, z)$,
with the planets located at $(r_{pi}, \phi_{pi}, 0)$ and the star located at
the origin of the coordinate system, the gravitational potential is
$\Phi_G = \Phi_c + \sum_{i=1}^N\Phi_{pi} ,$
where
\be \Phi_c = -{GM_* \over  r}  \ee
and 
\be
\Phi_{pi} = -{Gm_{pi}\over \sqrt{r^2 + r_{pi}^2 -2r r_{pi}
\cos(\phi-\phi_{pi})+ b^2}}.\ee

\noindent
Here, as in the papers PN2003, NP2003, PNS2004 and NP2004,
we have neglected the dependence
of the gravitational potentials on $z$
along with the vertical stratification of the disk, for reasons of computational
speed.
Thus the simulations are of cylindrical disks
(e.g. Armitage 1998, 2001; Hawley 2000, 2001; 
Steinacker \& Papaloizou 2002).
To model the effects of the
reduction of the planet potential with
vertical height, we have incorporated a softening length $b$ in the
potential.

\noindent We  use a locally isothermal equation of state in the form
\begin{equation}
P=c_s(r)^2 \cdot \rho,
\label{eq-state}
\end{equation}
with $c_s$ denoting the sound speed
which is specified as a fixed function of the cylindrical
radius $r.$
Expressions~(\ref{cont}) -- (\ref{eq-state}) give the basic 
equations used for the simulations.

\subsection{Planet Orbital Evolution}
In this work we allow the planet orbits to evolve due to the gravitational
forces
they experience from the disk as well as the central star.
For a single simulation we consider the evolution of
either six or three planets concurrently,
located at different positions within the disk.
These planets do not interact with each other gravitationally,
but affect each others evolution indirectly through perturbations they
make to the disk structure. In the case of low mass planets, the
effects of the turbulence will be of considerably greater
significance.

\noindent The equation of motion for planet $i$ is:
\be
\frac{d {\bf v}_{pi}}{dt} = -\frac{G M_*}{r_{pi}^3} 
{\bf r}_{pi} + {\bf f}_{di}
- 2 \Omega_f {\hat {\bf z}} \times {\bf v}_{pi} + \Omega_f {\hat {\bf z}}
\times (\Omega_f {\hat {\bf z}} \times {\bf r}_{pi} ).
\ee
The acceleration due to the disk is given by
\be
{\bf f}_{di} = - G \int_V \frac{ \rho({\bf r}) ({\bf r}_{pi}-{\bf r}) }
{(r^2 + r_{pi}^2 -2r r_{pi} \cos(\phi-\phi_{pi})+ b^2)^{3/2}} dV
\ee
where the integral is performed over the disk volume.
Note that the planet potentials are also cylindrical in that there
is no vertical component of gravity.
During the simulations we monitor the torque per unit mass experienced
by the planets, which is defined by
\be
{\bf T}_i = {\bf r}_{pi} \times {\bf f}_{di}.
\label{torque}
\ee
Note that the torque per unit mass is independent of the planet mass,
and so may be calculated for zero--mass `planetesimals' and 
for finite mass planets.

The calculations are performed in a uniformly rotating frame
with angular velocity $\Omega_f$. The planet equations of motion are
evolved using a simple leap-frog integrator, with correct centering
employed for inclusion of the Corioli's force.

\section{Numerical Method}
\label{num_method}
The numerical scheme that we employ is
based on a spatially second--order accurate method that computes the
advection using the monotonic transport algorithm (Van Leer 1977).
The MHD section of the code uses the
method of characteristics
constrained transport (MOCCT) as outlined in Hawley \& Stone (1995)
and implemented in the ZEUS code.
The code has been developed from a version
of NIRVANA written originally by U. Ziegler
(Ziegler \& Yorke 1997). 

\section{Computational Units}
\label{units}
We use units in which the central mass $M_*=1$, the gravitational constant 
$G=1$, and the radius $R=1$ corresponds to the radial location
of the inner boundary of the computational domain. The unit of time is 
$\Omega^{-1}=\sqrt{G M_*/R^3}$, although 
we report our results in units of the orbital period
at the disk inner edge, which we denote as $P(1)=2 \pi \Omega^{-1}$. 
Planets are positioned in the disk between radii $r_{pi}=2.2$ -- 3.2.
We note that the orbital period
at $r_{pi}=2.2$ is $P(2.2)=3.26P(1)$ and at $r_{pi}=3.2$ we have 
$P(3.2)=5.72P(1)$. Simulation runs times are typically $\simeq 500$ $P(1)$,
corresponding to $\sim 153 P(2.2)$ and $\sim 90 P(3.2)$.
If we take the computational radius $r=2.5$ to correspond to 5 AU,
then a simulation run time of $500 P(1)$ corresponds to $\sim 1335$ yr,
for a solar mass central star.

\section{Initial and Boundary Conditions}
\label{Init-bound}
\begin{figure}
\epsfig{file=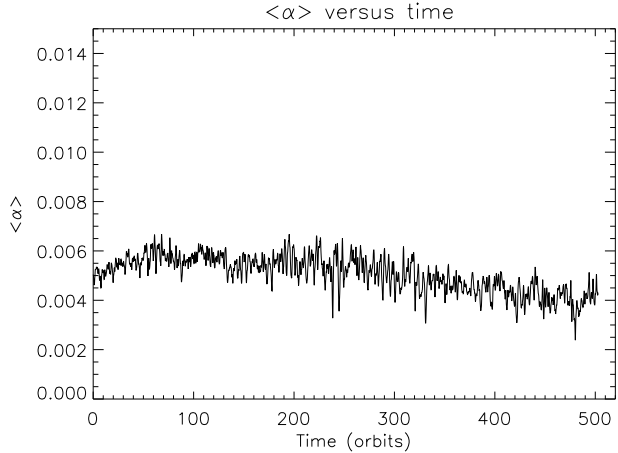, width=\columnwidth}
\caption{\label{fig1} This figure shows the time evolution of the
volume averaged total $\alpha$ value in the disk. Note that the time is
expressed in units of the orbital period at the inner edge of the disk.}
\end{figure}

The simulations presented here all use the same underlying
turbulent disk model. This model has a constant aspect ratio
$H/r \equiv c(r)/(r \Omega)=0.07$, where $\Omega$ is the disk angular
velocity measured in the inertial frame. The inner radial boundary of
the computational
domain is at $R_{in}=1$ and the outer boundary is at $R_{out}=5$.
The simulations were performed in the 
rotating frame with $\Omega_f=0.28668$,
which is the Keplerian angular velocity at a radius of $r=2.3$.

The boundary conditions employed are very similar to those described
in PN2003. Regions of the disk in the vicinity of the inner
and outer boundaries were given non--Keplerian angular velocity profiles
(uniform in the simulations described here)
that are stable to the MRI, and which have large values of the
density in order to maintain radial hydrostatic equilibrium. These regions
act as buffer zones that prevent the penetration of magnetic field
to the radial boundaries, thus maintaining the initial value of
net magnetic flux in the
computational domain. The inner buffer zone runs from 
$r=1.2$ to $r=R_{in}$.
The outer buffer zone runs from $r=4.5$ to $r=R_{out}$.

The creation of a relaxed turbulent disk model in which planets could be
immersed to examine their orbital evolution was
constructed using a multi--stage procedure. This is because
the initial relaxation of
a turbulent disk model can significantly modify
the initial surface density structure, due to radial
and temporal variations in the magnetic stresses during the relaxation. It is
desirable to minimise this effect so that a reasonably smooth profile
is used for the disk in which the planets are
inserted. Also, we wish to obtain a turbulent disk model
in which the volume averaged $\alpha$ value is $\simeq 5 \times 10^{-3}$,
as expected for protostellar disks, and the magnetic field strength in
the disk was adjusted to achieve this.

\begin{figure}
\epsfig{file=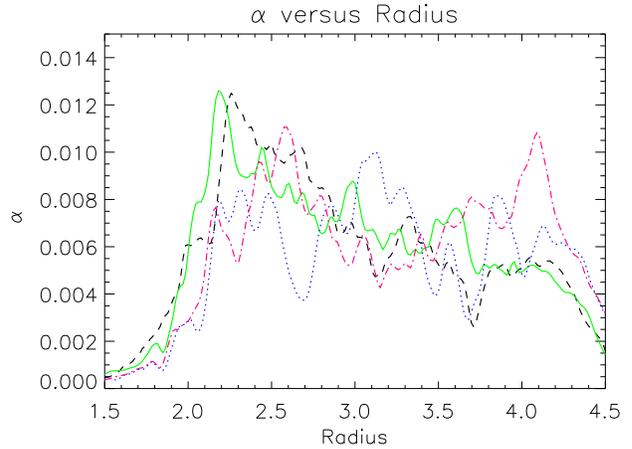, width=\columnwidth}
\caption{\label{fig2} This figure shows the evolution of the
radial variation in the total $\alpha$ value in the disk.
The solid line corresponds to a time of 15.6 orbits at the inner
disk edge, the dashed line corresponds to 31.4 orbits, the dotted line
to 71.2 orbits, and the dash--dotted line to 188.1 orbits.}
\end{figure}

The calculations presented in NP2004 used an initial
setup where $R_{in}=1$ and $R_{out}=8$. The numerical resolution
used was greater than that used here ([$N_r$, $N_{\phi}$, $N_z$]=
[450, 1092, 44], as opposed to [264, 608, 44] used in this paper),
but those simulations
could only be run for $\sim 25$ planet orbits. The higher resolution
used in NP2004 meant that zero net flux magnetic fields
could be used, requiring a self--sustaining dynamo to maintain the field.
The lower resolution adopted for the
simulations in this paper leads to the requirement that
net flux magnetic fields be used to sustain the turbulence over
long time periods. This is because the lower resolution simulations
cannot maintain an
active turbulent state for zero net flux fields. The use of net flux fields 
guarantees the continued existence of magnetic field
in the simulation domain, which helps maintain the turbulence.
A modest magnetic flux is used to
obtain a disk with modest turbulent stresses.

In order to obtain the desired disk model,
the following procedure was adopted:\\
({\it i}).  
The disk was initiated with a
toroidal magnetic field in a finite annulus in the disk between the
radii $1.5 \le r \le 4$. The initial field varied as
$B_{\phi}(r)=B_0/r$, and $B_0$ was defined such that the
initial ratio of volume integrated magnetic pressure to
thermal pressure was $1/\langle \beta \rangle=2.73 \times 10^{-3}$
(where $\beta=8 \pi P/B^2$).
The initial density distribution in the disk was $\rho(r)=\rho_0/r$
away from the buffer zones,
with $\rho_0$ defined such that the disk model would contain approximately
0.06 M$_{\odot}$ interior to 40 AU (assuming $r=2.5$ corresponds to
5 AU). In other words, the basic model is chosen
to be around three times more massive than the minimum mass solar nebula model.
The disk model was relaxed for a period of 191.7 orbits at
the disk inner boundary. \\
({\it ii}). The disk model was restarted at this point with
the initial density distribution re--established, but with all
other quantities remaining the same after the disk was relaxed for 191.7
orbits. The model was then run for a further 58.3 orbits. \\
({\it iii}). The model was restarted with the initial density
distribution re--established, and with the magnetic field
strength being reduced throughout the disk by a factor of
$1/\sqrt{2}$. The model was run for a further 37.2 orbits. \\
({\it iv}). The model was restarted after a further reduction in
the magnetic field strength by a factor of $1/\sqrt{2}$, and with
the initial density profile having been re--established.
The model was relaxed for a  further 32 orbits.
The final model has a volume averaged $\langle \alpha \rangle = 5.04 \times 10^{-3}$, and $1 / \langle \beta \rangle =0.012$, which are similar to
those obtained in the models presented in PN2003, PNS2004 and NP2004.

The disk model has a full $2 \pi$ azimuthal domain,
and vertical domain from $z_{min}=-0.14$ to $z_{max}=0.14$. 
Periodic boundary conditions were employed in the vertical
and azimuthal directions.

Gravitational softening of the protoplanet was employed in all
simulations, with the softening
parameter $b=0.33H$ and $H$ being evaluated at each planet position.  

\section{Simulation Results}
\label{results}
\begin{figure*}
\epsfig{file=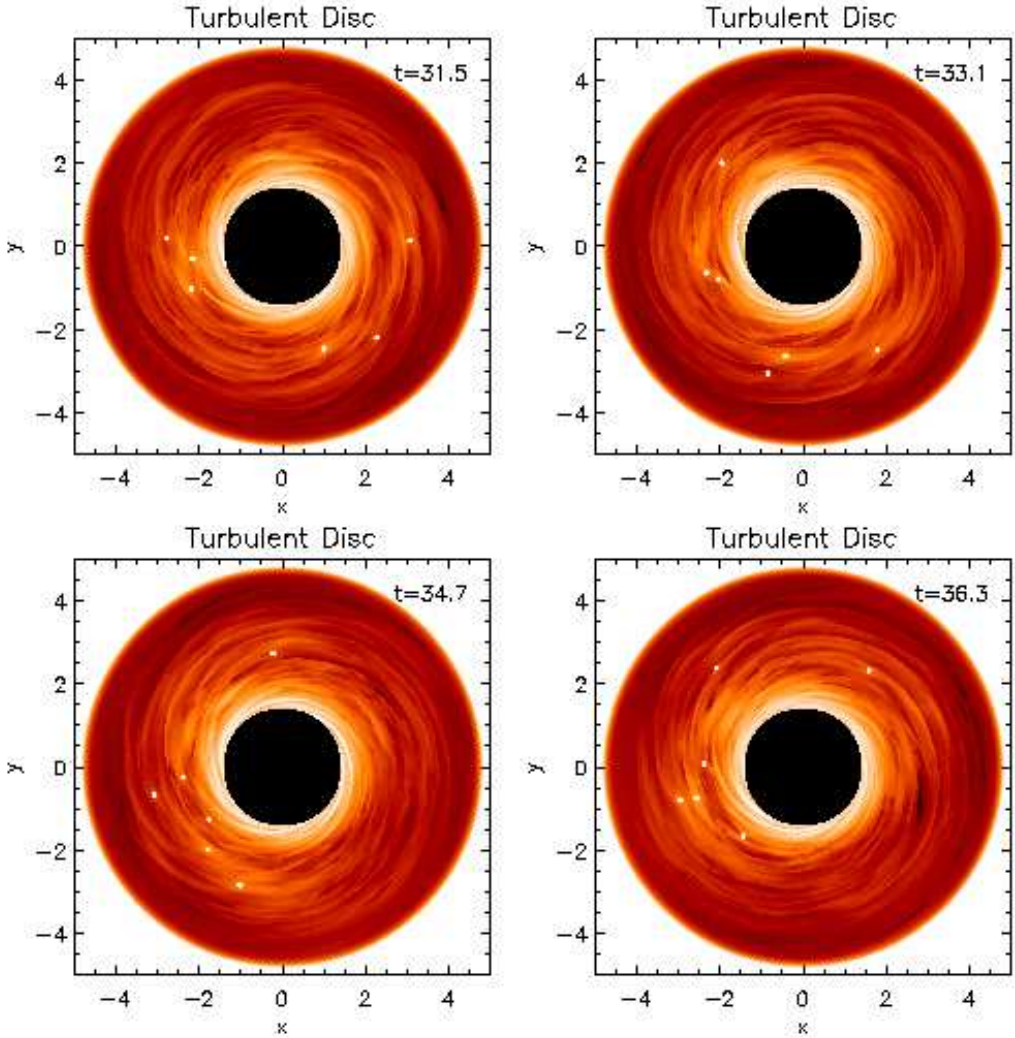, width=14cm}
\caption{\label{fig3} This figure shows snapshot images of the disk
midplane density for the run with $m_p=10$ M$_{\oplus}$ planets.
It is clear that the turbulent density fluctuations are typically larger
than the spiral wakes generated by the planets in this case. The times
corresponding to each image are shown at the top right of each panel,
in units of the orbital period at the disk inner edge.}
\end{figure*}

\subsection{Turbulent stresses}
\label{stresses}
In order to describe average properties of the turbulent models, we
use quantities that are both
vertically and azimuthally averaged over the $(\phi, z)$ domain
(e.g. Hawley 2000).
The vertical and azimuthal average of
$Q$ is defined through
\begin{equation}
{\overline {Q(r,t)}} ={\int \rho  Q dz d\phi \over \int  \rho dz d\phi}.
\end{equation}
The average is taken over the full $2\pi$ in azimuth.
The disk surface density is given by
\begin{equation}
\Sigma = {1\over 2\pi}\int \rho dz d\phi.
\end{equation}

\noindent  The  vertically and azimuthally
averaged Maxwell and
Reynolds stresses are defined as follows:
\begin{equation}
T_M(r,t)=2\pi
\Sigma{\overline{\left({B_r(r,\phi,z,t) B_\phi(r,\phi,z,t) \over 4\pi\rho}\right)}}
\end{equation}
and
\begin{equation}
T_{Re}(r,t)=2\pi
\Sigma
{\overline{\delta v_r(r,\phi,z,t)\delta v_\phi(r,\phi,z,t)}}.
\end{equation}
Here the velocity fluctuations $\delta v_r$ and $\delta v_\phi$
are defined through,
\begin{equation}
\delta v_r(r,\phi,z,t)=v_r(r,\phi,z,t)-{\overline{v_r}}(r,t),
\end{equation}
\begin{equation}
\delta v_\phi(r,\phi,z,t)=v_\phi(r,\phi,z,t)- {\overline{v_{\phi}}}(r,t).
\end{equation}
The Shakura \& Sunyaev (1973)
$\alpha$ stress parameter appropriate to the
total stress is  defined by
\begin{equation}
\alpha(r,t)=\frac{T_{Re}-T_M}{2\pi
\Sigma{\overline{ \left(P/\rho\right)}}},
\label{alphaeqn}
\end{equation}

\subsection{Disk Model}
The method used to set up the turbulent disk model is described
in section~\ref{Init-bound}. The subsequent evolution of the 
volume averaged $\alpha$ value 
is shown in figure~\ref{fig1}, for the duration
of the simulation performed with zero--mass planets described below.
Values of the volume averaged stress parameter
$\langle \alpha \rangle \simeq 5 \times 10^{-3}$ are sustained
throughout the duration of the simulation, which is approximately 500 orbits
at the disk inner edge. Snapshots of the radial distribution of 
$\alpha$ are shown in figure~\ref{fig2}, showing that a magnetically
active disk is sustained, at least for radii beyond $r \ge 2$.

Figure~\ref{fig3} shows snapshots of the midplane density for
the model with six 10 M$_{\oplus}$ protoplanets embedded.
These images show that the density perturbations induced by planets
of this mass and lower are exceeded by those generated by the
turbulence. The consequences of this on the planet orbital
evolution are described below for planet masses ranging between
$0 \le m_p \le 30$ M$_{\oplus}$. 
The effects of the turbulence on 
protoplanet orbital evolution are essentially stochastic, and for
this reason the simulations were performed with six or three
planets in order to sample the range of possible outcomes.

\subsection{Zero mass protoplanets -- `planetesimals'}
\label{mp0}
\begin{figure*}
\epsfig{file=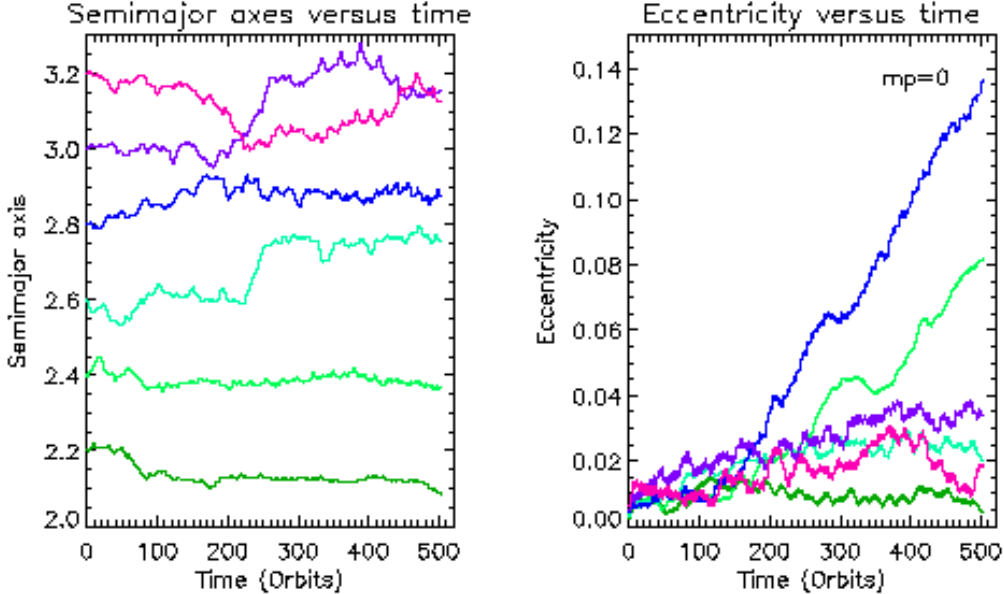, width=14cm}
\caption{\label{fig4} This figure shows the evolution of the planet
semimajor axes (left panel) and eccentricities (right panel) for the run 
with planet mass set to zero. It is clear from this plot that the time varying 
gravitational field of the disk causes stochastic migration of the
`planetesimals', and significant eccentricity growth.}
\end{figure*}

The orbital evolution of the six `zero--mass planets' is shown in 
figure~\ref{fig4}. These planets began with orbital radii
$r_{pi}=$ 2.2, 2.4, 2.6, 2.8, 3.0, 3.2 and randomised azimuthal positions.
From now on we shall refer to these zero--mass objects
as planetesimals, as they represent the evolution of bodies
whose gravity is too weak to substantially perturb the disk 
structure, and for which no underlying type I migration occurs.
We note that planetesimals experience gas drag, which is able to 
modify their orbits. We have neglected this effect here,
but models of planetesimal evolution 
including gas drag will be presented in a future paper.

The left hand panel of figure~\ref{fig4} shows the evolution of the 
planetesimals' semimajor axes, with each line representing the evolution of
a different planetesimal. It is clear from this plot that the gravitational
interaction
of the planetesimals with the turbulent density fluctuations causes the 
semimajor axes to change stochastically. This results in the planetesimals
undergoing a `random walk' in orbital location,
with variations of the semimajor axis at 
the 5 -- 10 percent level being observed over time periods corresponding to 
$\sim 100$ planetesimal orbits. 

The right hand panel shows the evolution of the planetesimal eccentricities.
Not surprisingly the effect of the turbulence is to excite 
significant orbital eccentricity in the planetesimal population.
Typical values obtained are $e \simeq 0.03$, but with values ranging from
$0.01 \le e \le 0.13$ over the run times of $\simeq 100$ planetesimal
orbits.

The results presented in figures~\ref{fig4}  
suggest that MHD turbulence will have a significant effect on the orbital
evolution of planetesimals in protoplanetary disks, which will in turn have
important repercussions for models of planet formation. In particular,
the effects of stochastic migration will help to prevent the isolation
of forming planetary embryos, and reduce the 
effects of orbital repulsion
and planetesimal shepherding during oligarchic growth
(Ida \& Makino 1993; Kokubo \& Ida 1998; Thommes et al 2003). 
This may allow the growth of 
planetary cores to proceed to the 10 - 15 M$_{\oplus}$ range, required
for the formation of giant planets via the core instability scenario 
(e.g. Pollack et al. 1996). However, it is also clear that 
MHD turbulence may hinder the growth of planetary cores by exciting
significant orbital eccentricities of the planetesimals, reducing the
effects of gravitational focusing. For planetesimals in the 1 - 10 km
range, it is unlikely that gas drag will be effective
in reducing these eccentricities because of the dominant contribution
from the turbulence as a source of excitation.
Under these conditions an alternative source of eccentricity damping may be
required, such as inelastic collisions between planetesimals, leading also
to fragmentation and more effective gas drag for the smaller fragments.
Of course, it then remains an open question how planetesimals
can form in the first place in such a turbulent environment. This may be 
related to the existence of `dead zones' in which the ionisation fraction
is too low to support MHD-driven turbulence (e.g. Gammie 1996).

\subsection{Finte mass planets}
The simulations presented in the next five 
sections (\ref{mp1} -- \ref{mp30}) were 
performed for either six or three planets placed in turbulent disks.
These planets do not interact with each other gravitationally,
as the purpose of these simulations is to sample the effects of turbulence
on single planets. The planets may, however, affect each other
indirectly through their influence on the disk. For planets whose 
orbits are well separated, this influence appears to be insignificant
compared to the turbulence. In some simulations, however,
there are planets that approach one another and occassionally cross
orbits, increasing the mutual indirect influence. 
Such simulations will no longer be sampling the effects of MHD 
turbulence on {\em isolated} planets. We note that this issue is likely to
be most important for the more massive planets. 

The planets in the 
simulation involving the 30 M$_{\oplus}$ objects 
do not approach each other.
The simulation including the 10 M$_{\oplus}$ protoplanets
does involve close encounters.
We have examined the torques experienced by the planets,
along with their semimajor axis and eccentricity evolution, 
and can detect no obvious sign that the evolution is being modified
by the close encounters. We are confident that the main results
of this paper are not contaminated by this effect.

\subsection{$m_p=1$ Earth mass protoplanets}
\label{mp1}
\begin{figure*}
\epsfig{file=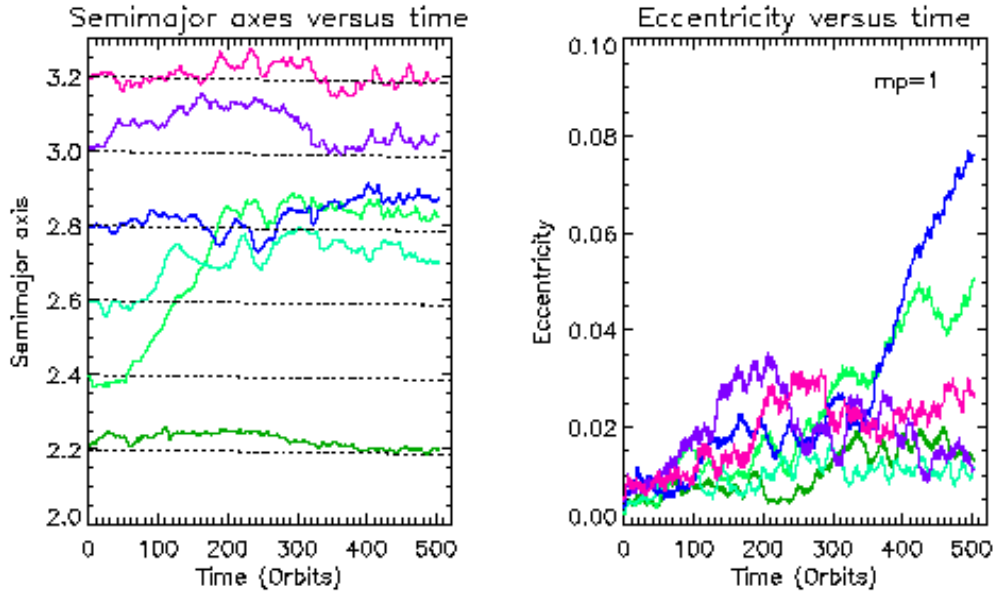, width=14cm}
\caption{\label{fig5} This figure shows the evolution of the planet
semimajor axes (left panel) and eccentricities (right panel) for the run
with $m_p=1$ M$_{\oplus}$. The dotted lines show the evolution of
planet orbits for simulations in which six 1 M$_{\oplus}$ planets were evolved
in a laminar disk model that was otherwise identical to the
turbulent disk model. The migration rates obtained are in quite good agreement
with the rates predicted by Papaloizou \& Larwood (2000) 
for planets with softened
potentials migrating in laminar disks.
The solid lines represent the planets
in the turbulent disk, and show that the planets undergo
migration similar to a random walk for the duration of the simulation,
with no clear tendency for the planets to migrate inward or outward.}
\end{figure*}
Figure~\ref{fig5} shows the orbital evolution of  the 1 M$_{\oplus}$ 
protoplanets. The dotted lines in the left panel show the evolution of
planet orbits for simulations in which six 1 M$_{\oplus}$ planets were evolved
in a laminar disk model that was otherwise identical to the
turbulent disk model. The solid lines, corresponding to planets in the
turbulent disk, show 
similar stochastic migration as observed for
the planetesimals. Indeed the protoplanet initially located at
$r_p=2.4$ migrates inward for $\sim 70$ orbits measured at the inner disk edge
at first, before migrating
outward for a sustained period of $\sim 130 $ orbits, modifying 
its semimajor axis by almost 20 percent during this time. The other planets
are also seen to migrate in a quasi--random fashion, 
but over less extreme distances.

The evolution of the eccentricities is plotted in the right hand panel.
A tendency for eccentricity growth to significant values for some
of the protoplanets is observed, in line with the results for the planetesimals.
It is interesting to observe however, that the peak eccentricities obtained
are somewhat less than those obtained in the planetesimal cases,
suggesting that the gravitational interaction with the disk
is providing eccentricity damping to counteract the effects of the turbulent
excitation.  This is a trend that continues as the protoplanet mass increases,
as described in the following sections.

\subsection{$m_p=3$ Earth mass protoplanets}
\label{mp3}

\begin{figure*}
\epsfig{file=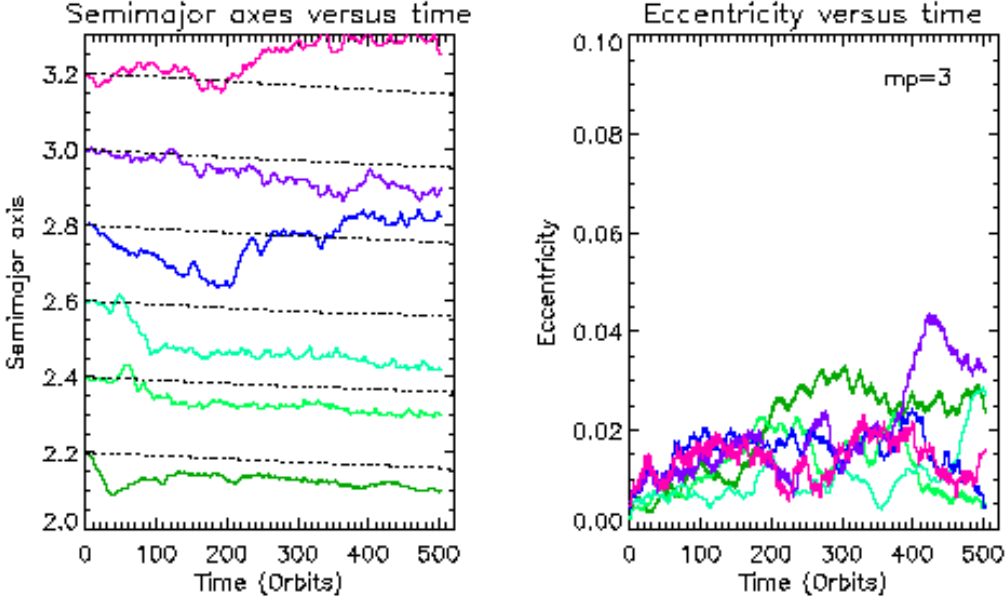, width=14cm}
\caption{\label{fig6} This figure shows the evolution of the planet
semimajor axes (left panel) and eccentricities (right panel) for the run
with $m_p=3$ M$_{\oplus}$. The figures show that the planets undergo
migration similar to a random walk for the duration of the simulation,
with no clear tendency for the planets to migrate inward or outward.}
\end{figure*}

The orbital evolution of the 3 M$_{\oplus}$ protoplanets is illustrated 
in figure~\ref{fig6}. The left panel shows a similar evolution 
for the semimajor
axes as has been described for the planetesimals and 1 M$_{\oplus}$ 
protoplanets. The semimajor axes are again observed to undergo 
modification in a stochastic manner, with variations at the 5 percent level 
being observed. The dotted lines show the results of a simulation
in which six 3 M$_{\oplus}$ protoplanets were evolved in a laminar disk.
The monotonic inward migration of each of these planets is clearly discernible,
suggesting that the planets do not greatly influence each others interaction
with the disk. Inward migration in this case occurs close to the expected
type I rate. The simulated torques are smaller than those presented
in Papaloizou \& Larwood (2000) by a factor of 0.71.

The eccentricity evolution is illustrated in the right panel of
figure~\ref{fig6}. Significant variations in the eccentricities are again
observed, with peak values of $e \simeq 0.05$ being excited. It is
interesting to compare the evolution with the planetesimal and 1 M$_{\oplus}$
cases, as there is a clear trend toward obtaining lower eccentricities
with increasing planet mass. 
It is likely that the eccentricity evolution is determined
by a balance between stochastic forcing due to the turbulence, and 
eccentricity damping due to interaction with the disk at coorbital
Lindblad resonances
(Goldreich \& Tremaine 1980; Artymowicz 1993).
This point is examined and discussed further in 
section~\ref{stochastic_torques}.

\subsection{$m_p=5$ Earth mass protoplanets}
\label{mp5}
\begin{figure*}
\epsfig{file=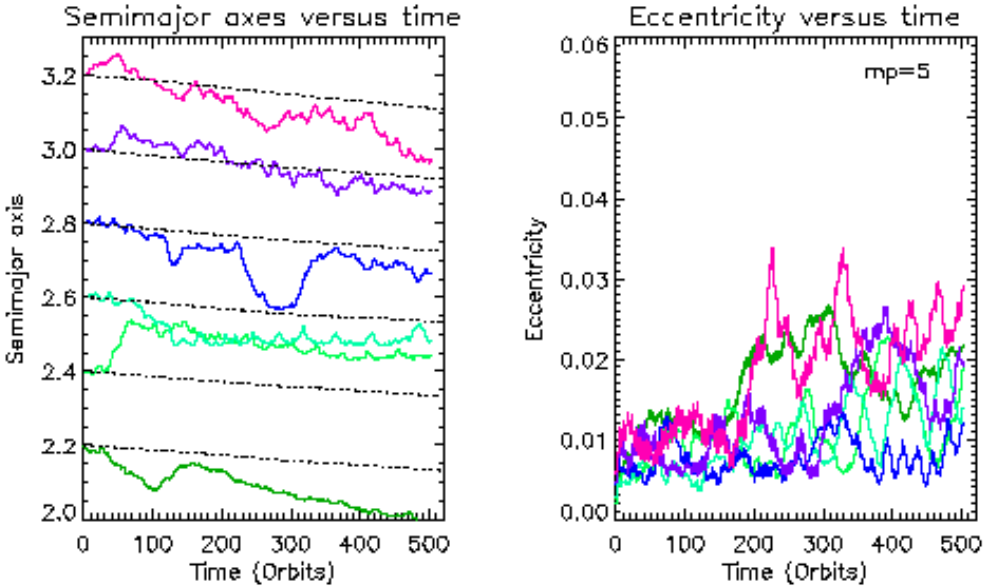, width=14cm}
\caption{\label{fig7} This figure shows the evolution of the planet
semimajor axes (left panel) and eccentricities (right panel) for the run
with $m_p=5$ M$_{\oplus}$. The figures show a general inward trend for the
migration, but this appears to be a statistical fluke as the $m_{p}=10$
M$_{\oplus}$ cases do not.
Note that the scale on the $y$ axis of the right panel has changed
compared with that in similar figures~\ref{fig4}--\ref{fig6}.}
\end{figure*}
The orbital evolution of the six 5 M$_{\oplus}$ protoplanets
is shown in figure~\ref{fig7}. The semimajor axes in the left panel
show a general inward trend, which might at first be thought to be
evidence that the usual type I migration is becoming dominant
over the stochastic forcing as the planet mass increases.
Closer inspection of the figures
shows that the protoplanets are still subject to substantial random forcing.
Furthermore the 10 M$_{\oplus}$ case discussed below suggests that
the general inward motion of the planets in figure~\ref{fig7} is probably
a statistical effect rather than evidence for type I migration overwhelming
stochastic forcing.
It is worth noting, however, that for all cases considered in which the 
planets have masses $m_{pi}> 1$ M$_{\oplus}$, the inner most planet
migrates inward at a rate similar to or greater that the expected type I
rate. This arises because the planet moves into a region of the disk
where it is less turbulent due to the initial set up, and where the
density increases during the course of the simulation due to disk accretion.

The eccentricity evolution shown in the right panel of figure~\ref{fig7} 
continues to show the trend of lower eccentricity for higher mass 
planets. Here the peak value for the eccentricity obtained is $e \simeq 0.04$,
but with eccentricities in the range $0.01 \le e \le 0.02$ being more typical.

\subsection{$m_p=10$ Earth mass protoplanets}
\label{mp10}
\begin{figure*}
\epsfig{file=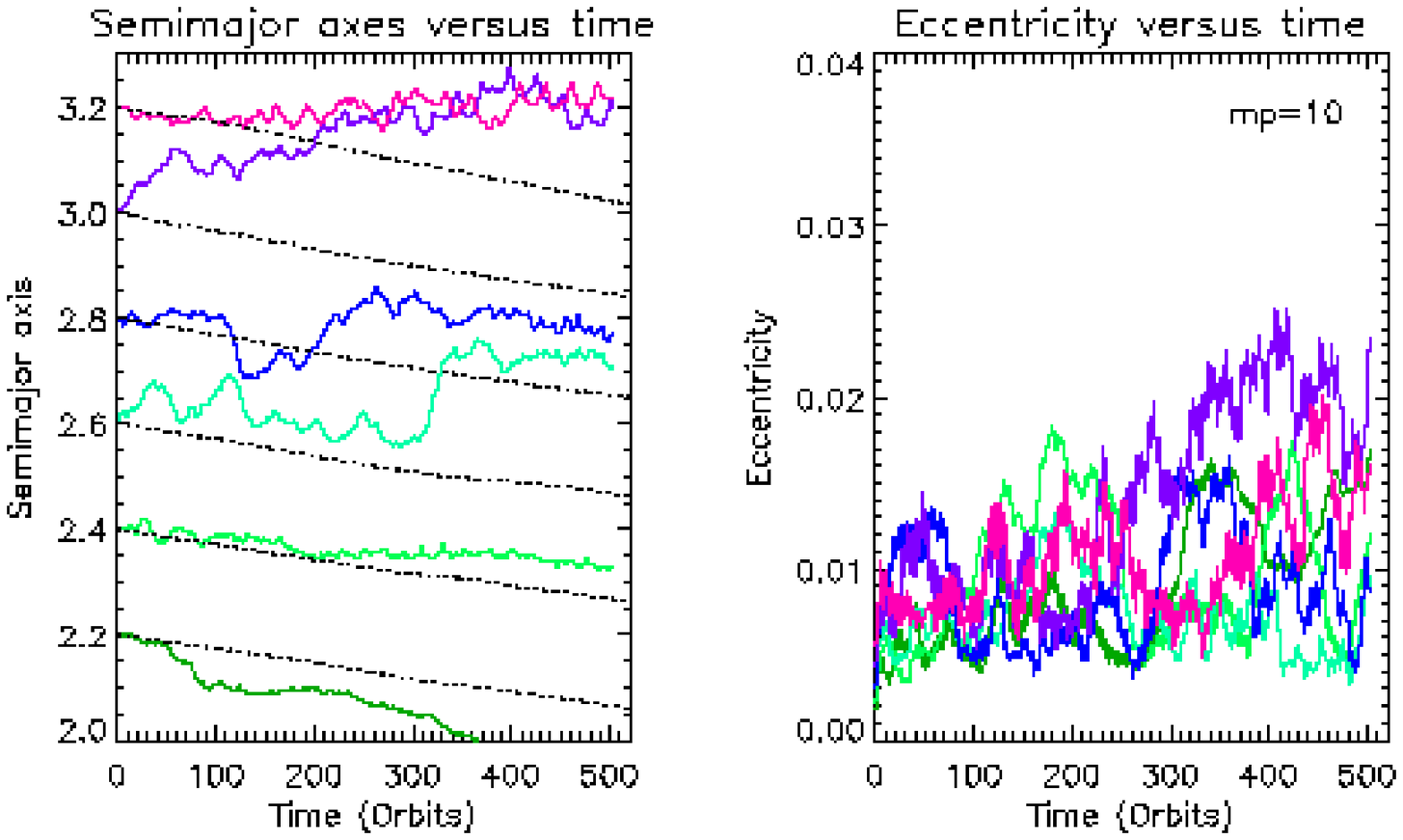, width=14cm}
\caption{\label{fig8} This figure shows the evolution of the planet
semimajor axes (left panel) and eccentricities (right panel) for the run
with $m_p=10$ M$_{\oplus}$. The figures show that the planets undergo
migration similar to a random walk for the duration of the simulation,
with no clear tendency for the planets to migrate inward or outward.
Note that the scale on the $y$ axis of the right panel has changed
compared with that in similar figures~\ref{fig4}--\ref{fig7}.}
\end{figure*}
Figure~\ref{fig8} shows the orbital evolution of the six 10 M$_{\oplus}$
protoplanets. Examination of the left panel shows that the planets'
orbital evolution remains dominated by stochastic forcing for this mass,
with the semimajor axes of three of the planets showing net increases for the
duration of the simulation. This suggests that the general inward motion
for the 5 M$_{\oplus}$ planets seen in figure~\ref{fig7} is a statistical
effect. The dotted lines in the left panel of figure~\ref{fig8} show
the evolution of six 10 M$_{\oplus}$ planets in a laminar disk.
The expected inward migration is observed for each planet
close to the predicted rate (Papaloizou \& Larwood 2000), 
showing that the planets have little effect on 
each others orbital evolution due to their perturbation of the disk.

The eccentricity evolution is shown in the right panel of figure~\ref{fig8}.
It is clear that the eccentricities remain quite small in this case,
with peak values reaching $e \simeq 0.025$. More generally, however,
the eccentricities remain at $\simeq 0.01$. The overall trend of eccentricity 
with  planet mass remains such that considerably less eccentricity growth
is observed for higher mass planets. The indication is that disk--planet
interaction at coorbital Lindblad resonances causes eccentricity damping,
and for planet masses
of $m_{pi} \sim 10$ M$_{\oplus}$ this dominates over turbulent forcing.

\subsection{$m_p=30$ Earth mass protoplanets}
\label{mp30}
\begin{figure*}
\epsfig{file=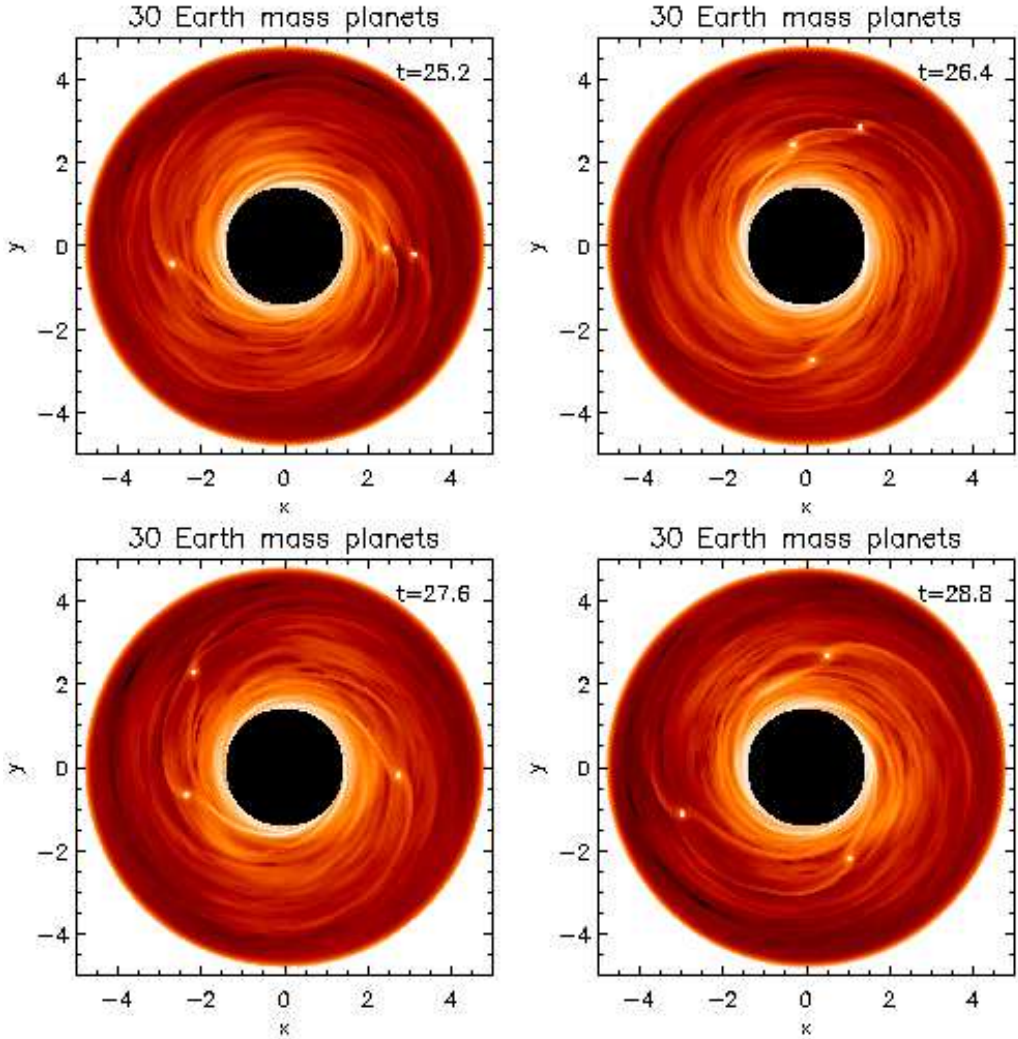, width=14cm}
\caption{\label{fig9} This figure shows snapshot images of the disk
midplane density for the run with $m_p=30$ M$_{\oplus}$ planets.
It is clear that the turbulent density fluctuations are of similar
amplitude to the spiral wakes
generated by the planets in this case. The times
corresponding to each image are shown at the top right of each panel,
in units of the orbital period at the disk inner edge.}
\end{figure*}
We consider the orbital evolution of three 30 M$_{\oplus}$ protoplanets,
in contrast to the six planets considered for smaller mass objects,
since they exert larger perturbations on the disk. Snapshots of the
midplane density distribution are shown in figure~\ref{fig9}, which
show that the planets are now massive enough to
excite density waves of similar amplitude to the background
turbulent fluctuations. A similar trend was noted by NP2004.

\begin{figure*}
\epsfig{file=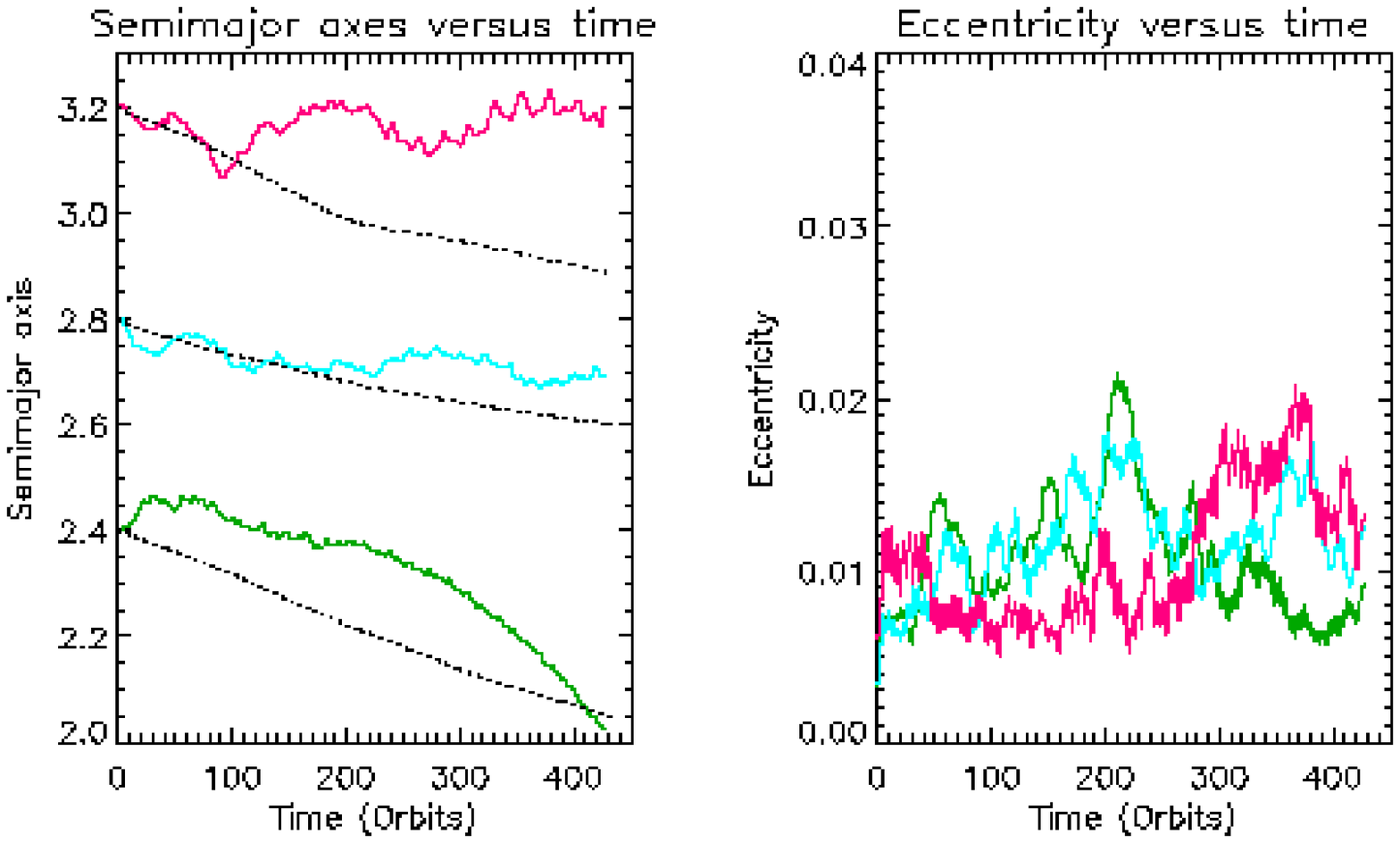, width=14cm}
\caption{\label{fig10} This figure shows the evolution of the planet
semimajor axes (left panel) and eccentricities (right panel) for the run
with $m_p=30$ M$_{\oplus}$. The figures show that the outer 
two planets undergo
migration that is substantially different to the laminar disk runs.
The inner planet migrates primarily inward as the disk turbulence is reduced
in the inner regions. Note that the scale on the $y$ axis in the right
panel differs from
that of similar plots pertaining to lower mass protoplanets.}
\end{figure*}

The semimajor axis evolution is shown in the left panel of figure~\ref{fig10}.
The inner most planet is observed to migrate inward, such that after 400 orbits
it has reached the same location as predicted for a laminar disk. This is again related to the fact that the inner disk is less turbulent.
The outer two planets, however, show a significant deviation from the
trajectories followed by planets in a laminar disk model. 
The middle planet undergoes noisy inward migration, but at a rate
substantially less than that obtained for the laminar model.
The outermost planet  undergoes essentially no net migration for the duration
of the model run time. It is clear that stochastic effects due to the
turbulence continue to have an impact on the migration of 30 M$_{\oplus}$ 
protoplanets.

The eccentricity evolution is shown in the right 
panel of figure~\ref{fig10}.
The values obtained are similar to those for the 10 M$_{\oplus}$ planets,
indicating little growth of the eccentricity because the damping induced
by the underlying type I resonant disk interaction.

\section{Stochastic torques and type I migration}
\label{stochastic_torques}

\begin{figure*}
\epsfig{file=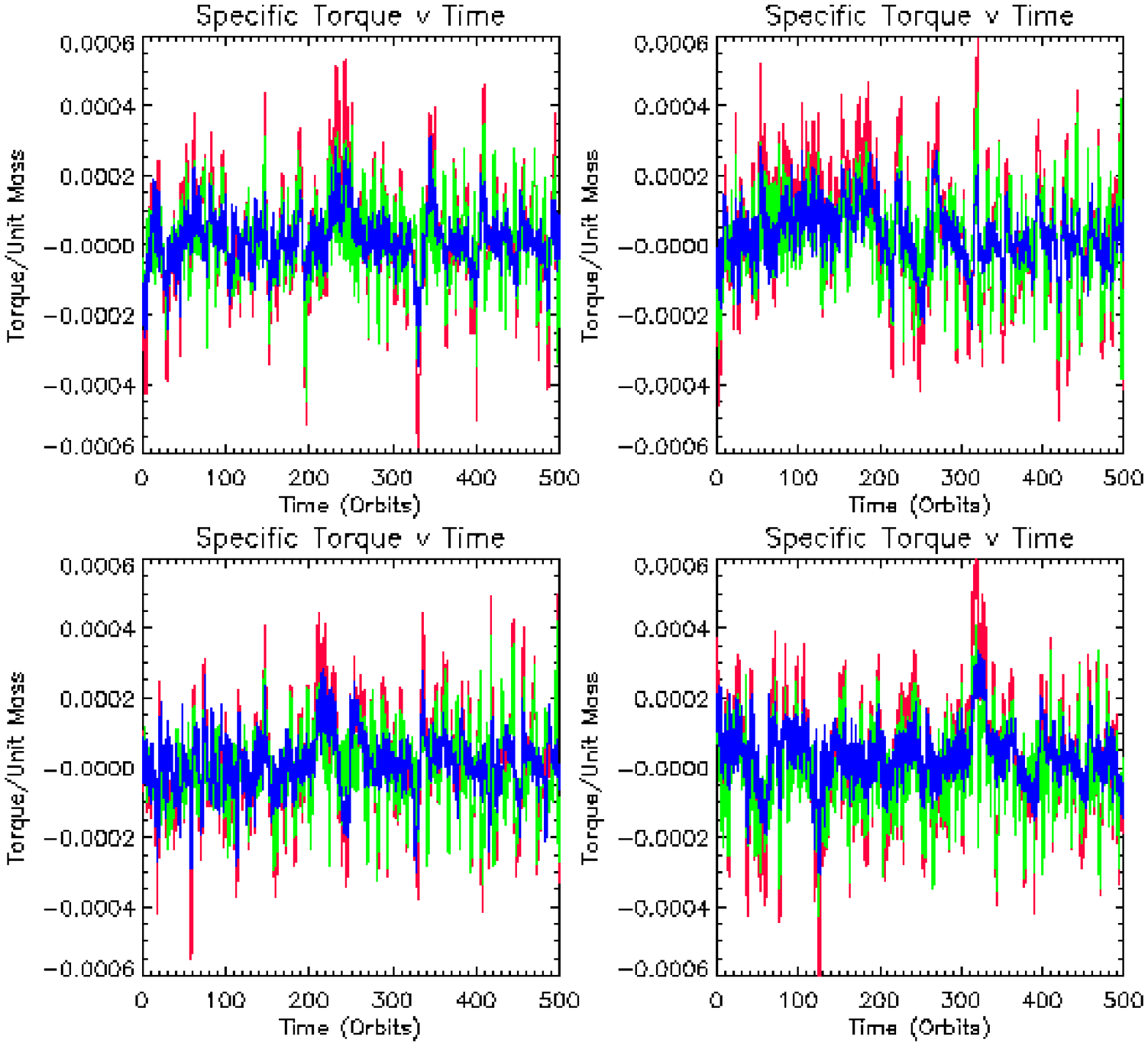, width=14cm}
\caption{\label{fig11} This figure shows the time evolution of the torques per
unit mass for four different planets selected from four different simulations.
Moving from left to right and from top to bottom the planet masses and 
initial orbital radii are:
($m_{pi}=0$, $r_{pi}=2.6$); ($m_{pi}=1$, $r_{pi}=2.4$);
($m_{pi}=3$, $r_{pi}=2.8$); ($m_{pi}=10$, $r_{pi}=2.6$)
The light grey (green)
line corresponds to the torque exerted
on the planet by the outer disk, the black (blue) line shows the
torque due to the inner
disk, and the dark grey (red) line shows the total torque. Moving from $m_p=0$
to $m_p=10$ M$_{\oplus}$ we observe that the 
inner and outer disk torques tend to separate,
but that the torque fluctuations due to the 
turbulence remain significantly
greater in amplitude than the mean torque.}
\end{figure*}

We now examine the torques experienced by the protoplanets during the
simulations in more detail. In view of the large number of planets
considered, we restrict our discussion to a few specific examples which
are illustrative of the range of behaviour observed in the simulations as a
whole. We also discuss the implications for type I migration of low
mass planets in turbulent disks.

\subsection{Time evolution of stochastic torques}
\label{time_evol}
Figure~\ref{fig11} shows four examples of the time evolution of the
torque per unit mass experienced by the planets in the simulations.
The torque per unit mass is defined by equation~(\ref{torque}), 
and is presented in the units described in section~\ref{units}.
Moving from left to right and from top to bottom, the panels show results
from simulations with: ($m_{pi}=0$, $r_{pi}=2.6$); ($m_{pi}=1$, $r_{pi}=2.4$);
($m_{pi}=3$, $r_{pi}=2.8$); ($m_{pi}=10$, $r_{pi}=2.6$), where
the radii, $r_{pi}$, refer to the initial orbital radii of the planets. 
In each panel, the torque on the planet due to the inner disk is shown by the
black (blue) line, that due to the outer disk is shown by the light grey 
(green) line, and
the dark grey (red) line represents the total torque. In all cases the torque is
a highly variable quantity, as discussed previously in NP2004.
On the same scale, the type I torque due to an equivalent laminar disk is
$\simeq 1.5 \times 10^{-6} \left(\frac{m_{pi}}{M_{\oplus}}\right)$. This
value was obtained from the simulations of low mass planets embedded
in laminar disks represented by the dotted lines in 
figures~\ref{fig5} -- \ref{fig8}.
Comparison with the migration time
given by equation~(32) of Papaloizou \& Larwood (2000) shows that our
simulated torques are smaller by a factor of $\simeq 0.71$ compared to their
fit to linear calculations.
Stochastic variations relative to type I torques therefore typically
range between 10 and 100 for planet masses in the range 1 to 10 M$_{\oplus}$,
but with peak fluctuations being up to four times larger than these values.
As was noted in NP2004, these large fluctuations make it difficult
to separate the torques due to the inner and outer disk when inspecting
figure~\ref{fig11} for planet masses $m_p \le 10$ M$_{\oplus}$.

\begin{figure*}
\epsfig{file=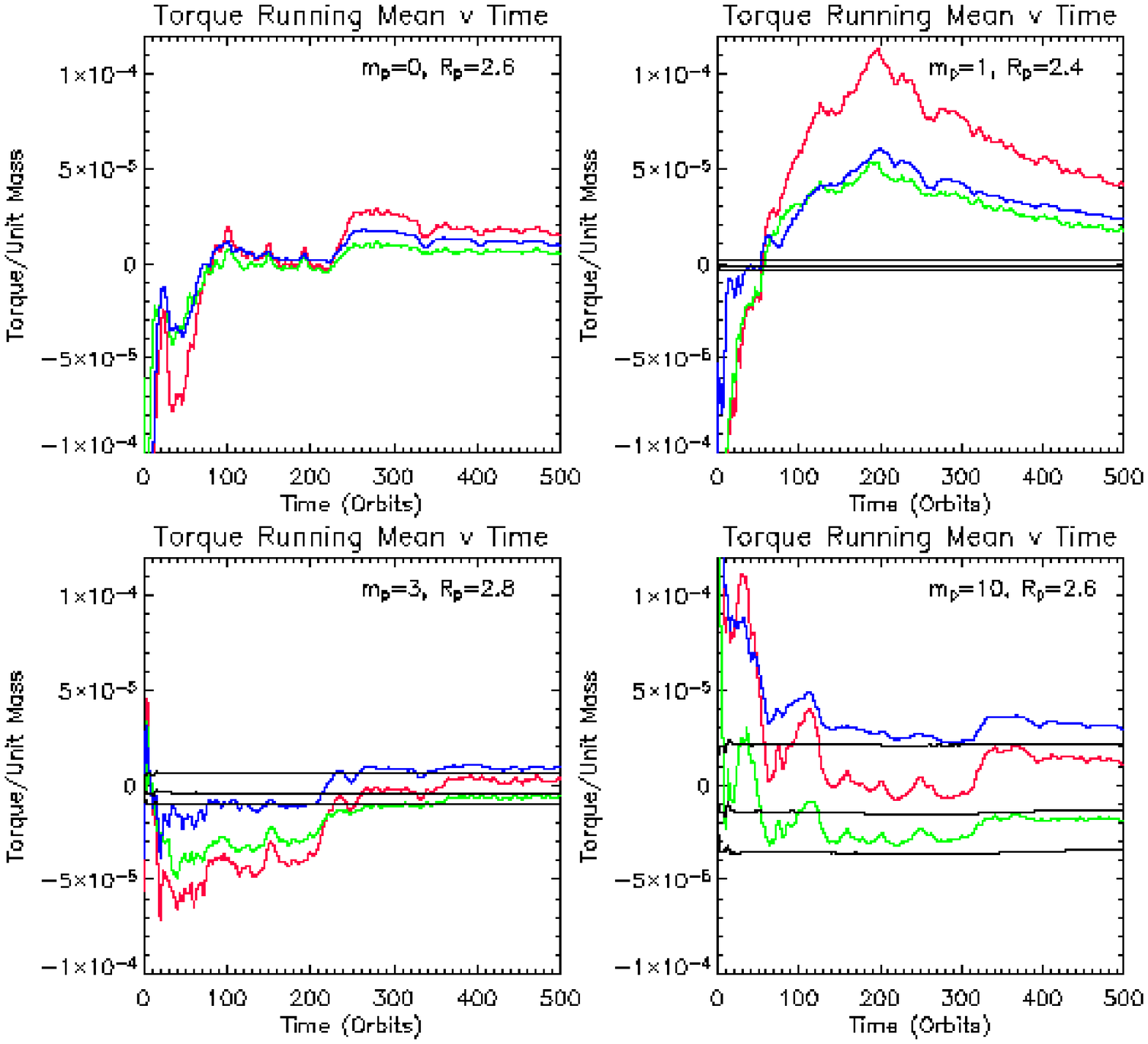, width=14cm}
\caption{\label{fig12} This figure shows the time evolution of the running
mean torques per
unit mass for the four simulations shown in figure~\ref{fig11}.
The planet masses and initial orbital radii are indicated at the top
right of each panel. The light grey (green)
line corresponds to the torque exerted
on the planet by the outer disk, the black (blue) line shows the
torque due to the inner
disk, and the dark grey (red) line shows the total torque. 
Also plotted with black lines are the running means obtained from 
equivalent simulations
with viscous, laminar disk models. Here the upper line is due to the inner disk,
the lower line is due to ther outer disk, and the middle line is the total
torque.}
\end{figure*}

The time evolution of the running means of the torques displayed in
figure~\ref{fig11} are shown in figure~\ref{fig12}, along with the 
running means of the torques from the equivalent laminar disk simulations.
In all cases, the running means do not converge to the equivalent type I
laminar disk torques during the simulations, as one would expect from
inspecting the migration histories presented in 
figures~\ref{fig4} -- \ref{fig8}. 
If good convergence of the running
means to the appropriate type I torque is possible, 
then these figures show that runs times
considerably longer than those presented here will be required to achieve it.
Such runs are currently not feasible computationally.

\begin{figure*}
\epsfig{file=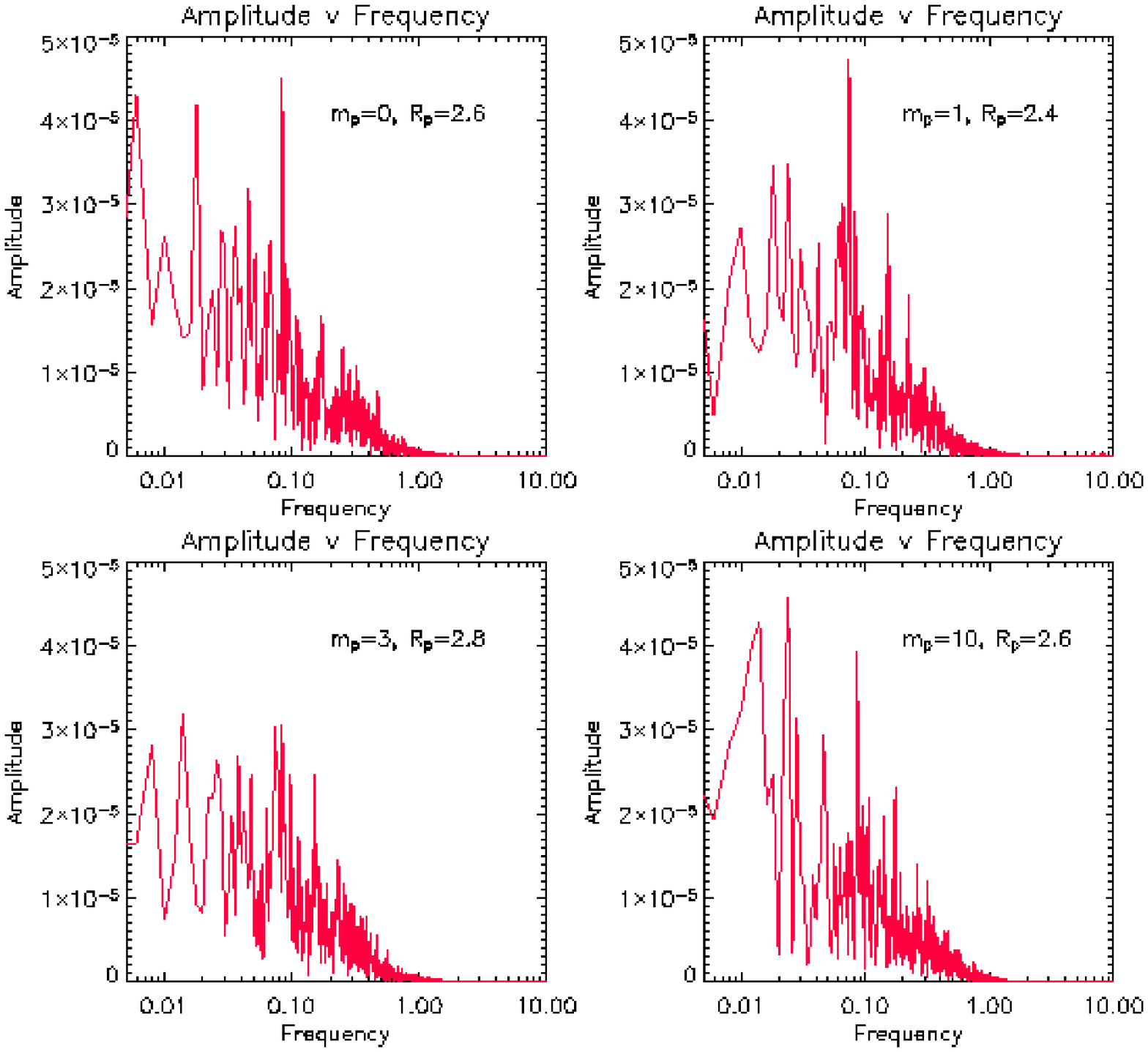, width=14cm}
\caption{\label{fig13} This shows the power spectrum of the total torques
plotted in figure~\ref{fig11}. The main point to be taken from these plots
is that significant power exists in frequencies that correspond to the
longest time scales in the simulation, suggesting that low frequency
variations in the torques experienced by the planets are important.}
\end{figure*}

This latter point concerning convergence 
is illuminated to some degree by figure~\ref{fig13}, which
shows the power spectrum of the temporal evolution of the total torques
plotted in figure~\ref{fig11}. In these figures, a frequency of between
0.2 -- 0.3 represents the orbital periods of protoplanets located at orbital
radii between  $r_{pi}=$2.23 -- 2.92. The power spectra are
obtained by Fourier analysis of the time evolution of the
total torques generated during the simulations. These torques are
output every ten time steps during the simulations, with the time
interval between each output being $\Delta t$. 
If $N$ is the total number of data points then we can define
discrete frequencies by
\begin{equation}
\omega_n=\frac{2 \pi n}{N \, \Delta t}
\label{freq}
\end{equation}
where $n=-N/2, ..., N/2$.
The discrete Fourier transform of the torques is then
\begin{equation}
H(\omega_n) = \frac{1}{N}\sum_{k=0}^{N-1} T_k \exp{(i \omega_n t_k)} 
\label{fourier}
\end{equation}
where $t_k$ is the time corresponding to data point $k$, and $T_k$ is the 
torque.
We define an amplitude by
\begin{equation}
A(\omega_n) = \sqrt{| H(\omega_n) |^2 + | H(-\omega_n) |^2}.
\label{amplitude}
\end{equation}
$A(\omega_n)$ is plotted against $|\omega_n|$ in figure~\ref{fig13}.

The power spectra show that there is significant
signal in the long term variation of the torques, with the total run time 
corresponding to a frequency of 0.002 in these plots.
This explains why the averaged torques 
in figure~\ref{fig12} do not converge to a well defined value: the stochastic
torques contain contributions with significant amplitude whose associated
time scale of variation is similar to the simulation run time.

The origin of these low frequency fluctuations is unclear.
One possibility is that global communication across the disk
on the viscous evolution time may lead to modification of the
local turbulence on that time scale. This may in turn feed into
the torque fluctuations experienced by embedded protoplanets.
The disks we consider in this paper do not have a very large range in
radii, but for the disk parameters $H/r=0.07$ and $\alpha=5 \times 10^{-3}$
the viscous communication time between the region where the planets
are located (centred around $r=2.8$) and the disk outer edge ($r=4.5$)
is $\tau_\nu \simeq 8800$ orbits. Given that the simulations have
been run for 500 orbits, communication between neighbouring regions
of the disk may be influencing the turbulence on this time scale.
One test of this would be to compare the power spectra of fluctuations in
global disk calculations with local simulations performed in the shearing
box. If the local simulations do not show the long term
fluctuations then this could be taken as evidence that global communication
is important. If low frequency fluctuations are observed,
then it would indicate that these fluctuations are
the property of the locally operating dynamo.

\begin{figure*}
\epsfig{file=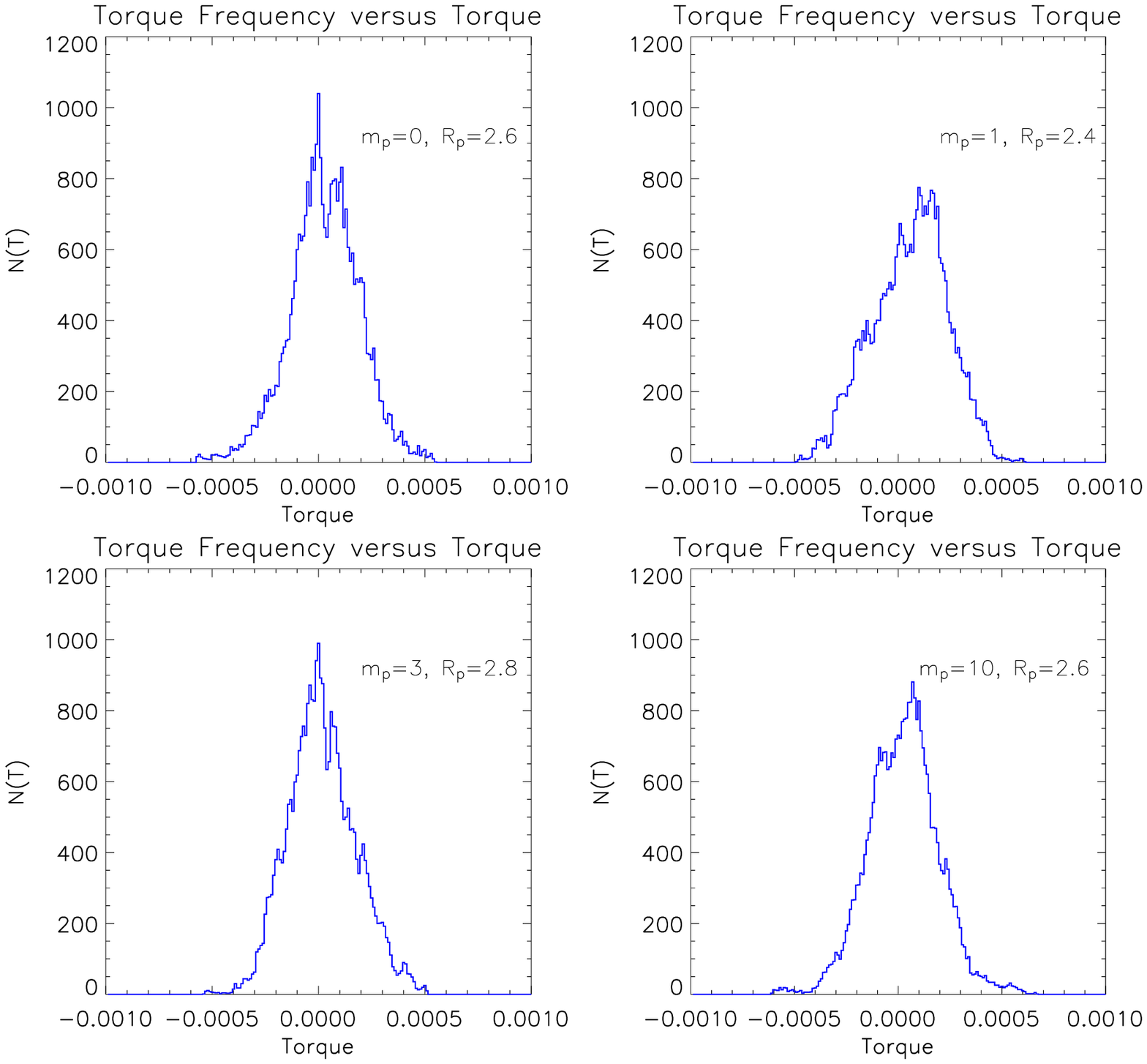, width=14cm}
\caption{\label{fig14} This shows the frequency distribution of the torques
plotted in figure~\ref{fig11}. The plots show that the stochastic
torques experienced by the planets have distributions that are fairly
close to being Gaussian with standard deviation 
$\sigma(T) \simeq$ 1-- 2 $\times 10^{-4}$.}
\end{figure*}

The torque frequency distributions corresponding to figure~\ref{fig11}
are plotted in figure~\ref{fig14}. These were calculated by sampling the
torque experienced by the planets every 10 time steps during the simulations.
The torques were then binned with bin widths $\Delta T= 10^{-5}$, 
and the number in each
bin counted. Although not strictly Gaussian,
these distributions are similar to Gaussian profiles with standard
deviation between $\sigma_T \simeq 1$ -- $2\times 10^{-4}$.

\subsection{Comparing stochastic migration and type I migration torques}
\label{stoch+typeI}
In a previous study, NP2004 considered the orbital evolution of low
mass planets in turbulent disks, and treated the issue of torque
convergence as a simple signal--to--noise problem.
The basic assumption here is that the torque experienced by the
planet, $T(t)$, consists of a linear superposition of a rapidly varying term,
$T_f(t)$, caused by turbulent fluctuations, and a constantly acting
type I torque $<T>$:
\begin{equation}
T(t)= <T> + T_f(t).
\label{signal-to-noise}
\end{equation}
Assuming that the stochastic torques are Gaussian distributed
(the Central Limit Theorem tells us that even if the torques are
not normally distributed, their cumulative effects will be similar
to normally distributed torques), the time average of 
equation~(\ref{signal-to-noise}) becomes
\begin{equation}
{\overline T} = <T> + \frac{\sigma_T}{\sqrt{t_{tot}}}.
\label{time-average}
\end{equation}
Here $\sigma_T$ is the standard deviation of the torque amplitude,
and $t_{tot}$ is the total time elapsed, measured in units of
the characteristic time for the torque amplitude
to vary. Convergence of the torques toward the underlying type I value
is expected to occur once the two terms on the right hand side become 
equal. 

Using a very simple by--eye inspection of `torque versus time' plots,
similar to figure~\ref{fig11}, NP2004 estimated the time for torque variation 
and the typical amplitude of the fluctuating torques. Using these in
equation~(\ref{time-average}) they
derived a time for
torque convergence of $\simeq 70$ planet orbits for a
10 M$_{\oplus}$ protoplanet. Following a similar procedure, inspecting a
blow--up of figure~\ref{fig12} leads to a by--eye  estimate of the
torque variation time $\simeq$ half a planet orbital period (the same as used 
by NP2004).
Inspecting the torque distrbutions in figure~\ref{fig14}
gives a value of the torque
standard deviation $\sigma_T \simeq 1.5 \times 10^{-4}$.
Inserting these values into equation~(\ref{time-average}), and
using a type I torque value of 
$1.5 \times 10^{-6} \left(\frac{m_{pl}}{M_{\oplus}}\right)$,
gives an estimated time for torque convergence equivalent to 50 planet
orbital periods when the protoplanet mass is 10 M$_{\oplus}$. 
The modest difference in this
estimate compared to NP2004 arises because the estimates of the standard
deviation of torque fluctuations differ.
For planets distributed in radius 
as they are in our simulations,
this translates into a convergence time of between 163 -- 286 orbits measured
at $r=1$. This crude analysis predicts that we should see inward
migration for the 10 M$_{\oplus}$ planets in figure~\ref{fig8},
which is clearly not the case.

Treating the issue of torque convergence using the above approach
has a number of problems.
The main one is illustrated by figure~\ref{fig13}, which clearly demonstrates
that there is no single frequency associated with the torque variation,
and furthermore indicates that a simple by-eye examination
of torque versus time plots can be misleading about which time scale
is dominant. A more sophisticated treatment should take account of
the fact the stochastic torques have significant amplitude distributed
across a broad range of frequencies. 
We have taken a different approach to addressing the issue of torque
convergence that ought to be more accurate.

In order to test whether the evolution of low mass planets
in turbulent disks can be described as a superposition of
stochastic torques and constantly acting type I torques,
we have performed a series of N-body simulations
in which particle orbits were evolved under the influence
of prescribed forces. These included stochastic forces,
which were calculated using a Fourier analysis of the migration torques
obtained in the MHD simulation with planet masses $m_{pi}=0$
(i.e. the `planetesimal' run shown in figure~\ref{fig4}). This
Fourier analysis allowed forces from the actual MHD simulations
to be incorporated into simple integrations of particle orbits.
The Fourier transform of the torques as a function of time
is given by equation~(\ref{fourier}) in section~\ref{time_evol}.
The reconstructed torque at time $t$ 
is then given by the inverse transform:
\begin{equation}
T(t)= \sum_{n=0}^{N-1} H(\omega_n) \exp{(-i \omega_n t)}
\label{inverse_fourier}
\end{equation}
This expression can be calculated for any value of $t$ required, and
so provides a means of including the stochastic torques
in an N-body simulation. We note, however, that the torques
calculated during the MHD simulations were only output
once every ten time steps, rather than at every time step.
This means that our reconstructed
torques are not exactly the same as the torques experienced by the planets
during the MHD simulations. The variation of the torque experienced
by a planet over ten time steps is not great 
(the torque shows significant variation over $\sim$ one thousand time steps),
so we expect the reconstructed torques to be a good approximation
to the real torques.

In addition to including stochastic force,
type I torques taken from equation~(32) of Papaloizou \&
Larwood (2000) were included, for which the inward migration
rate is linearly proportional to the planet mass. We note that
comparing torques from laminar disk simulations,
and those given in Papaloizou \& Larwood (2000), show that
those obtained in the simulations are smaller by a factor of $\simeq 0.71$, 
so we multiplied the type I torques
used in the N-body simulations by this factor.

The particle orbits were integrated using
a simple leap--frog integrator. The time step used was held constant,
and was chosen to be equal to the mean time step size which arises from the
Courant condition in the MHD simulations. The time step calculated
during the MHD simulations varied by only a few percent 
throughout the simulations.

\begin{figure*}
\epsfig{file=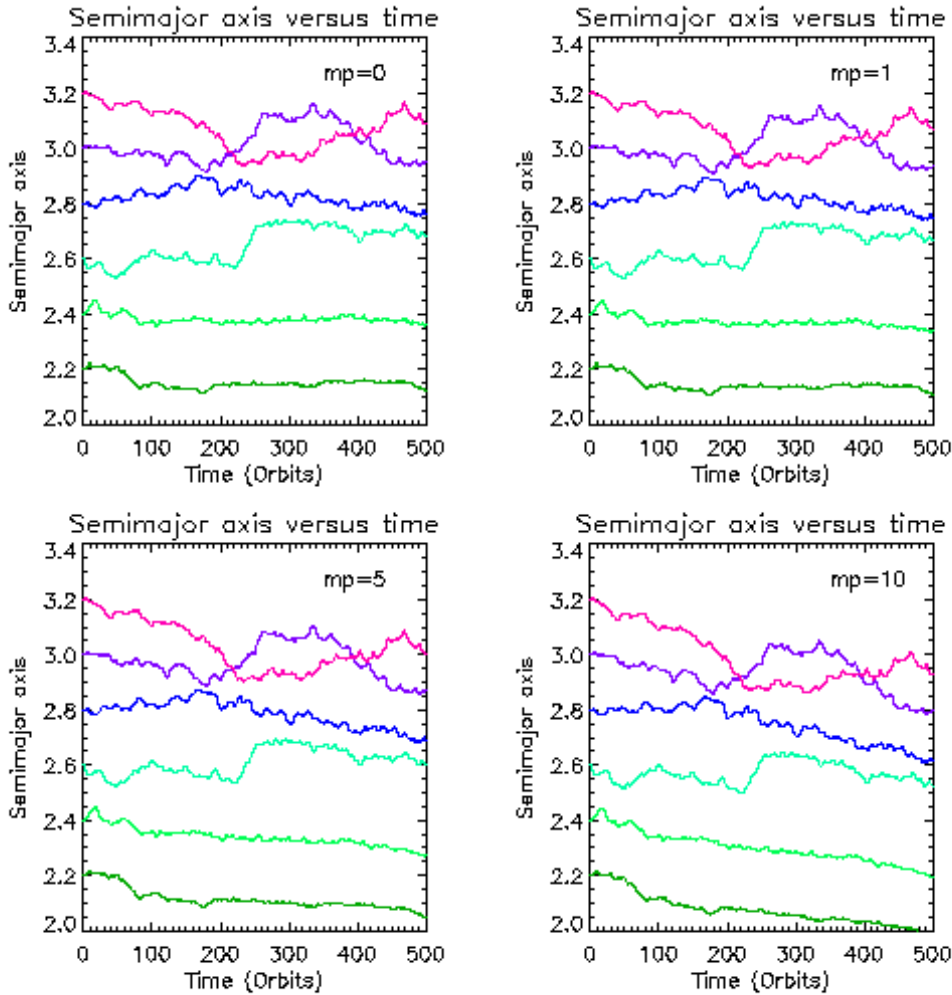, width=14cm}
\caption{\label{fig15} This figure shows the evolution of semimajor
axes from simulations
in which the stochastic torques generated by the MHD simulation
with $m_p=0$ were reconstructed using
Fourier analysis, and type I migration torques were included using 
a scaled version of equation~(32) from Papaloizou \& Larwood (2000). 
For planet masses 
lower than 10 M$_{\oplus}$,
stochastic evolution of the orbits is dominant. 
For $m_p=10$ M$_{\oplus}$,
the effects of the superposed type I migration torques are
dominant.}
\end{figure*}

The results of these runs are shown in figure~\ref{fig15},
where the planet mass is
indicated in the upper right corner of each panel.
For zero--mass planets,
orbit evolution similar to that shown in figure~\ref{fig4}
was obtained. This indicates that the sampling of torques after
every ten time steps in the MHD simulations leads to
reconstructed torques that are quite accurate.
For successively higher masses, these orbit trajectories
were modified by having a small inward drift superposed.
For planet masses $m_{pi} \ge 10$ M$_{\oplus}$, a definite inward
drift that overwhelms the stochastic migration can be observed.
It is worth noting that the inward drift observed
for the 10 M$_{\oplus}$ cases in figure~\ref{fig15} is not matched
by the actual MHD simulation presented in figure~\ref{fig8}.

We should point out that we have only sampled six realisations of the
turbulence when performing the calculations presented in figure~\ref{fig15}.
On this basis alone it would be wrong to draw too strong a conclusion about the
role of underlying type I torques in the MHD simulations.
Inspection of figures~\ref{fig4}--\ref{fig8}, and figures~\ref{fig13} and
\ref{fig14} show that the torque distributions can differ substantially from
run to run, such that the results shown in figure~\ref{fig15} may simply
be a reflection of a minority of possible outcomes. 

To test this further we ought to perform additional simulations of
MHD turbulence with zero--mass test particles being used to sample
the torque fluctuations. This, however, would be computationally expensive.
The remaining simulations we have presented, for
finite mass planets, contain the effects of turbulent fluctuations
{\em and} spiral density waves 
(not always visible against the turbulent backdrop) 
excited by the planets. As such they cannot be used to sample cleanly the
role of turbulence--induced fluctuations, but including them in the
current analysis is still useful. 

A 10 M$_{\oplus}$ planet in a laminar disk migrates a distance of
$\Delta \, r \simeq 0.15$ in 500 orbits (as shown by the dotted lines in 
figure~\ref{fig8}). We now consider the effect of superposing
such a drift on the
migration histories of all planets  plotted in figures~\ref{fig4} --
\ref{fig8} whose orbits started between radii $r_{pi}=2.6$--3.2.
Of this sample of twenty planets,
a subset of eleven show net outward migration prior
to the superposition. Of these eleven, only one would show net outward
migration after superposition (this would
be the 10 M$_{\oplus}$ planet that was initiated at $r_{pi}=3.0$), two
would show no net migration, and eight would show a net inward drift.
Based on this apparent tendency for the simulations as a whole to
show inward migration when type I drift is superposed, which 
differs from the outcome observed in figure~\ref{fig8} for which the planet
mass is 10 M$_{\oplus}$, there is a hint that the effects of type I
migration may be diminished in a turbulent disk. This may arise
because of density and pressure perturbations in the vicinity of
the planet modifying the excitation of spiral density waves and the
usual bias between inner and outer disk.
At present, however,
the results of the simulations are sufficiently ambiguous for such
a conclusion to be premature. It remains possible that the fluctuation spectrum
associated with the orbital evolution shown in figure~\ref{fig8}
is such that it is simply able to overcome the effects of type I migration
for the duration of the simulations due to significant contributions from
low frequency components. Resolution of these questions requires
simulations to be run for longer duration, and for customised simulations
to be performed that examine in detail the exchange of angular momentum
between planet and turbulent disk. These calculations will be the subject of
future publications.

\begin{figure*}
\epsfig{file=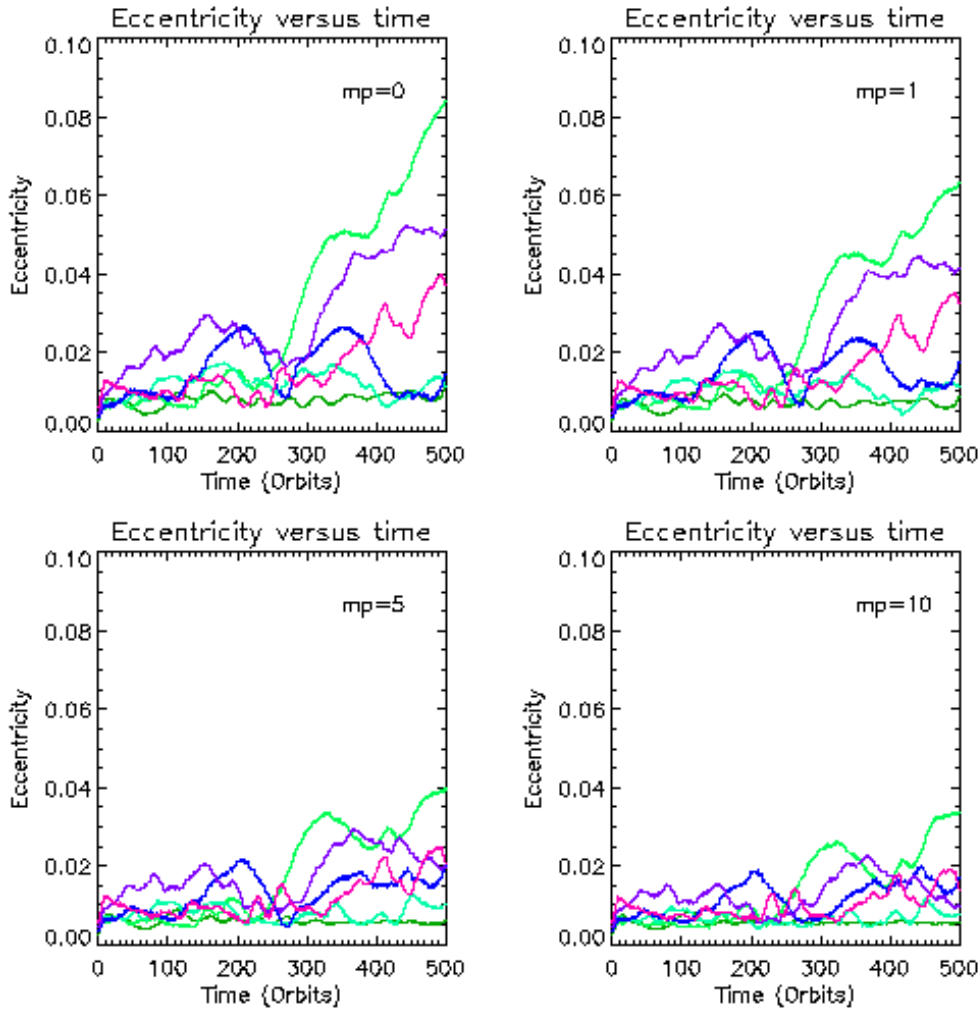, width=14cm}
\caption{\label{fig16} This figure shows the evolution of eccentricities
from simulations
in which the stochastic torques generated by the MHD simulation
with $m_p=0$ were reconstructed using
Fourier analysis,
and eccentricity damping was
included using equation~(38) from Papaloizou \& Larwood (2000).
The planet mass in each case is indicated
in the top right of each panel. A clear trend in decreasing eccentricity
may be observed as the masses of the planets increase, in broad agreement with
the full MHD simulation results presented in sections~\ref{mp0}--\ref{mp30}.}
\end{figure*}

\section{Eccentricity Evolution}
\label{eccentricity}
In order to examine the origin of the behaviour of the orbital
eccentricities observed in figures~\ref{fig4} -- \ref{fig8}, we performed
simulations similar to those described at the end of section~\ref{stoch+typeI}.
Particle orbits were evolved using a simple leap--frog integrator,
and with prescribed forces included representing stochastic
eccentricity forcing and eccentricity damping
associated with interaction at coorbital Lindblad resonances.
The stochastic forces were
calculated using a Fourier reconstruction of the radial and azimuthal forces
experienced by the `planetesimals' in the MHD simulation for which
the planet mass $m_{pi}=0$, results for which are shown in figure~\ref{fig4}.
The eccentricity damping was included by using equation~(38)
of Papaloizou \& Larwood (2000). As already described, 
the source of this damping is
primarily interaction with the disk at coorbital Lindblad resonances
(e.g. Artymowicz 1993). 

We calculated orbital evolution for
planets with masses $m_{pi}=0$, 1, 3, 5, 10, and 30 M$_{\oplus}$ as in the
MHD simulations. The masses simply control the degree of eccentricity
damping, whereas the eccentricity driving is independent of mass.
A selection of the resulting eccentricity behaviours
are shown in figure~\ref{fig16}. These figures show that good agreement is
obtained with the evolutionary trends
found in the full MHD simulations, suggesting that
the eccentricity evolution obtained in those runs was indeed due to
stochastic forcing by the turbulence, and 
damping due to resonant disk interaction.
It is interesting to note that the agreement with the eccentricity evolution 
is better than that obtained when considering the superposition of
type I torques and stochastic forces in section~\ref{stoch+typeI}.
The reason for this is simply that the type I migration torques
rely on there being an asymmetry between the
inner and outer disk contributions, which may not occur as
the density field near the planet varies erratically.
The eccentricity damping, however, is largely produced by coorbital
Lindblad torques located near the orbit of the planet.
These should always be present throughout the MHD simulations, and operate as a
constant source of damping, although the rate of damping may vary with time 
due to fluctuations
induced in the disk structure by the turbulence.

\section{Discussion}
\label{discussion}
\subsection{Long term evolution of planetesimal orbits}
\label{longterm}
Here we consider only objects for which gravitational 
fluctuations due to the turbulent disk dominate the dynamics
rather than gas--drag forces. Plantesimals with radius 1 -- 10 km
fall into this category.
The migration histories of the simulated planetesimals plotted in 
figure~\ref{fig4} show that such bodies will undergo
substantial migration {\em via} a random walk if
protoplanetary disks are globally turbulent.
The mean migration distance of the planetesimals
in figure~\ref{fig4} is $\Delta \, r \simeq 0.09$
over a time of 500 orbits measured at $r=1$.
If we make the simplifying assumption that
the migration distance scales according to a random walk, so
that $\Delta \, r \propto \sqrt{t}$, where $t$ is the time
elapsed, and further take the location $r=2.5$ in the simulations
to be equivalent to  5 AU, then the time taken for a typical
planetesimal to migrate a distance equal to its semimajor
axis is $t \simeq 1$ Myr. The stochastic torques
were generated by a disk model that is approximately
three times more massive than a minimum mass model.
A lower mass disk would generate a correspondingly longer
migration time.

It is clear that planetesimals in turbulent disks will undergo significant
migration during the lifetime of the nebula. From the point of view
of forming planets quickly, turbulence--induced changes to the orbital
elements can have both a positive and a negative effect. Increased
mobility generally acts to favour forming planets more quickly.
One of the major issues facing the theory of
planet formation is forming the cores of the giant planets fast enough
to accrete gas before dispersal of the nebula. Although estimates
of the core formation time based on models of runaway growth suggest
relatively short core formation times (e.g. Pollack et al 1996), 
simulations indicate that runaway growth does not continue all the way
to the completion of core formation.
Runaway growth slows down and enters
a stage of `oligarchic growth' when the eccentricities and inclinations
of the planetesimals are pumped up by the larger embryos (Ida \& Makino 1993;
Kokubo \& Ida 1998).
Simulations by Thommes et al (2003) indicate that core formation becomes 
difficult, in part due to gap formation in the planetesimal disk by the 
larger embryos and in part due to the dynamical
excitation of the planetesimal disk,
leading to cores of $\sim$ few M$_{\oplus}$ forming in a few million years.
The
isolation of cores may be alleviated by the increased mobility of embryos and
planetesimals in a stochastic migration scenario. In addition, the existence
of regions of the disk where the turbulence is weaker than in others
may lead to the accumulation of planetesimals there that could speed up
accumulation processes.

The eccentricity
and (and presumably) inclination driving by turbulence indicated 
by the right panel of figure~\ref{fig4}, however, 
may slow down the early stages of planet formation. 
In particular the runaway growth of planetesimals will be affected,
as the velocity dispersion of small bodies
needs to be smaller than the escape
velocity from the larger accreting objects. Indeed a large velocity dispersion 
generated by turbulence may lead to destructive rather than accumulative
collisions between planetesimals, leading to questions about how 
planetesimals form at all in a turbulent disk. It should also
be noted that the large scale migration of 
icy planetesimals from the outer solar system into the inner solar
system may lead to the inner planets being more enriched in volatiles
than is observed. 

These problems may be alleviated
if the stochastic torque values obtained in the MHD simulations performed
here, using cylindrical disk models, turn out to be overestimates when models
including vertical stratification are computed. 
A vertical gradient may then occur
in the relative density fluctuations due to magnetic buoyancy effects 
(e.g. Stone et al 1996; Stone \& Miller 2000), which reduce the
torque fluctuations at the disk midplane. The effects of a vertical gradient in
the ionisation fraction (e.g. Gammie 1996) would act in a similar manner.

\subsection{Long term evolution of protoplanet orbits}
\label{longterm-planets}

The situation regarding the long term evolution of
planetesimals in turbulent disks is clear -- they will 
slowly diffuse throughout the disk. The situation regarding high
mass, gap forming planets also seems to be clear. They will migrate
inward at the effective viscous evolution of time of the disk,
as in the standard type II migration picture
(Nelson \& Papaloizou 2003, 2004). The situation regarding low 
mass protoplanets remains ambiguous.
The central question remains: can stochastic forces overcome
type I migration and prevent at least some low mass
protoplanets from falling into the star prior to accreting a
gaseous envelope and becoming gas--giants ?

The answer to this question depends on how effective type I torques
are in a turbulent disk, and whether the disk
turbulence can generate fluctuations with the required 
amplitudes and temporal behaviour to counterbalance
the inward drift on planet formation time scales of
$\sim 1$ Myr. The simulations presented in sections~\ref{mp0} -- \ref{mp30}
and section~\ref{stoch+typeI}
indicate that type I torques may be diminished in turbulent disks
compared with those that arise in laminar disks, but the evidence
is far from conclusive. The power spectra presented in figure~\ref{fig13}
show that the disk turbulence is capable of generating long term
fluctuations with the appropriate amplitude to significantly affect type I
migration, but there is no way of predicting whether
such fluctuations can persist for time scales appropriate to
planet formation itself. This question can only be answered by simulations
whose run times are currently too long to be performed.

The origin of the longer time scale fluctuations observed in the simulations
remains unknown. We have suggested that global communication across the disk
on the viscous time scale may feed into the fluctuations, contributing
to their long term temporal evolution. The viscous communication time
from the outer edge to the planet forming region in our model disks is
$\simeq 8000$ orbits. For naturally occuring disks, that form through the
collapse of molecular clouds and have radii in excess of 100 AU, the viscous
evolution time exceeds 1 Myr. It is not unreasonable to suppose that 
the nature of the turbulence  in the planet forming region between 
1 -- 10 AU may be affected by global communication through the disk
on this time scale. This may in turn feed into the torques
experienced by forming protoplanets.

Other effects that may arise in naturally occuring disks are the formation
of surface density depressions or `gaps'. Simulations show that variations 
in the magnetic stresses occur radially throughout the disk, and persistence
of these can lead to the formation of `gaps' (e.g. Hawley 2000; Steinacker \&
Papaloizou 2002). Such features may play an important role in
modifying the migration of low mass planets. The tendency of
MHD turbulence to assist in gap formation for more massive planets
has already been noted by Nelson \& Papaloizou (2003) and
Winters, Balbus \& Hawley (2003).
Radial variations 
in the strength of the disk turbulence may also play a role.
The existence of `dead zones' has been suggested, particularly in
the planet forming region around 5 AU (e.g. Gammie 1996; 
Fromang, Terquem \& Balbus 2002), with an inner active zone maintained
through thermal ionisation of alkalii metals.
A dominant outward flux of waves excited
by a more active interior disk may provide a bias that helps
stochastic torques counter balance  inward type I migration.

\section{Conclusions}
\label{summary}
In this paper we have presented simulations of low mass 
protoplanets embedded in turbulent, magnetised disks. The planet orbits
are evolved, with forces due to gravitational interaction with the disk 
being included. The disk models are cylindrical, as the vertical component
of gravity is neglected, but are otherwise designed to be similar
to disks thought to be present around young stars during the planet
formation epoch (i.e. disk thickness
$H/r=0.07$; disk viscosity $\alpha \sim 5 \times 10^{-3}$).

Planet masses, $m_{p}$, ranging between zero and 
30 M$_{\oplus}$ were considered.
The $m_p=0$ `planetesimals' were found to migrate through the disk
in a manner similar to a random walk, and in some cases their
eccentricity was found to increase to values $e \simeq 0.1$. The random walk
element of their evolution provides a means of maintaining significant
mobility of the planetesimal swarm during planet formation. This may help
reduce the problems of orbital repulsion and core isolation found during 
simulations of giant planet core formation (e.g. Thommes et al. 2003).
However, the increase in eccentricity may present a problem during the
runaway growth stage of planet formation, which requires a dynamically cold
disk of planetesimals (e.g. Wetherill \& Stewart 1993). Further
simulations that include gas drag and vertically stratified disk models
are required to obtain a more accurate picture of this eccentricity growth.

The protoplanets with finite masses that were studied also showed 
evidence of stochastic migration and eccentricity growth.
The eccentricity evolution was such that the protoplanets
with mass $m_p=1$ and 3 M$_{\oplus}$ showed peak eccentricities of
$e \simeq 0.08$ and $0.045$, respectively, whereas the $m_p=10$ and 30 
M$_{\oplus}$ planets had peak eccentricities of $e \simeq 0.02$.
This is in broad agreement with expectations, as the interaction of
low mass protoplanets with disks causes eccentricity
damping due to interaction with material at coorbital Lindblad resonances
(e.g. Artymowicz 1993). This damping is expected to scale linearly with
the protoplanet mass, leading to the observed behaviour.

For all protoplanet masses in the range $1 \le m_p \le 30$ M$_{\oplus}$,
there was clear evidence of stochastic torques being dominant over the type I
migration expected to occur in laminar disks. We considered an ensemble of
models, and while some planets showed a definite trend toward inward migration,
there were examples for each mass considered for which
stochastic migration was the dominant effect. This behaviour arises
in part because the turbulent fluctuations contain low frequency
components whose amplitudes are such that they can
significantly modify type I migration over the simulation
run times that are currently feasible. In addition, we analysed
the effect of superposing type I migration expected for  laminar disks
onto the stochastic migration obtained in turbulent disks.
This analysis suggested that inward migration ought to be more
apparent in the MHD simulations than was found to be the case,
providing a hint that type I migration may be modified in turbulent disks.
We note, however, that the simulations are not conclusive on this latter point.

A number of important issues remain to be addressed. These include
a determination of the role of gas drag in modulating the effects
of eccentricity growth of planetesimals, and the effects of vertical
disk structure in influencing the strength of the stochastic torques
at the disk midplane. The question of whether stochastic torques
can prevent type I migration of protoplanets over planet formation
time scales still remains to be settled, and can be partially
answered by performing longer simulations. The potential
modification of type I migration in turbulent disks can be 
addressed by customised simulations that examine the exchange
of angular momentum between disk and planet
These will be the subjects of future publications. 

\section{Acknowledgements}

The computations reported here were performed using the UK Astrophysical
Fluids Facility (UKAFF) and the QMUL HPC facility purchased under the
SRIF initiative. I would like to thank Richard Frewin, Alex Martin, and David
Burgess for assistance with the computations presented here.
Comments received from an anonymous referee have led to significant
improvements in this paper.

\end{document}